\documentclass[11pt]{article}
\usepackage{amsmath,amssymb,amsthm,color,epsfig,mathrsfs}
\usepackage{amsbsy,placeins}
\usepackage{sectsty}

\allsectionsfont{\bfseries\centering}

\newtheorem{theorem}{Theorem}

\newtheorem{prop}[theorem]{Proposition}

\newcounter{spslist}

\newcounter{geqncount}
    {\refstepcounter{equation}%
     \setcounter{geqncount}{\value{equation}}%
     \setcounter{equation}{0}%
  }%
    {\setcounter{equation}{\value{geqncount}}}

\newtheorem{Theorem}{Theorem}[section]

\newcommand{\R}{{\cal R}}
\renewcommand{\P}{{P}}
\newcommand{\pa}{\hat{p}}
\newcommand{\pb}{\hat{\bar{p}}}

\renewcommand{\Re}{\text{Re}\,}
\renewcommand{\Im}{\text{Im}\,}
\newcommand{\half}{{\textstyle{\frac{1}{2}}}}

\begin{document}

\bibliographystyle{plain}

\renewcommand{\thefootnote}{}

\begin{center}
{\bf \Large An Exactly Solvable Model for Nonlinear Resonant Scattering\footnote{This work was supported by the National Science Foundation under grants DMS-0807325 (SPS) and DMS-0707488 (SV) while the authors were hosted at the Universidad Carlos III de Madrid in the fall of 2010.  We thank the NSF and the Universidad Carlos III for their support.}
}
\end{center}

\vspace{0.2ex}

\begin{center}
{\scshape \large Stephen P. Shipman$^\dagger$ and Stephanos Venakides$^\ddagger$} \\
\vspace{1ex}
{\itshape $^\dagger$Department of Mathematics, Louisiana State University; \\ Lockett Hall 303; Baton Rouge, LA \ 70803, USA; \ shipman@math.lsu.edu}\\
{\itshape $^\ddagger$Department of Mathematics, Duke University; \\
Box 90320; Durham, NC \ 27708, USA; \ ven@math.duke.edu}
\end{center}

\vspace{3ex}
\centerline{\parbox{0.9\textwidth}{
{\bf Abstract.}\
This work analyzes the effects of cubic nonlinearities on certain resonant scattering anomalies associated with the dissolution of an embedded eigenvalue of a linear scattering system.
These sharp peak-dip anomalies in the frequency domain are often called Fano resonances.
We study a simple model that incorporates the essential features of this kind of resonance.  It features a linear scatterer attached to a transmission line with a point-mass defect and coupled to a nonlinear oscillator.  We prove two power laws in the small coupling ($\gamma\to0$) and small nonlinearity ($\mu\to0$) regime.  The asymptotic relation $\mu\sim C\gamma^4$ characterizes the emergence of a small frequency interval of triple harmonic solutions near the resonant frequency of the oscillator.  As the nonlinearity grows or the coupling diminishes, this interval widens and, at the relation $\mu\sim C\gamma^2$, merges with another evolving frequency interval of triple harmonic solutions that extends to infinity.  Our model allows rigorous computation of stability in the small $\mu$ and $\gamma$ limit.  In the regime of triple harmonic solutions, those with largest and smallest response of the oscillator are linearly stable and the solution with intermediate response is unstable.
}}

\vspace{3ex}
\noindent
\begin{mbox}
\end{mbox} \\
\begin{mbox}
\end{mbox}
\vspace{3ex}

\hrule
\vspace{1.1ex}

\section{Introduction of the nonlinear model} 

\begin{figure}
\centerline{
\scalebox{0.47}{\includegraphics{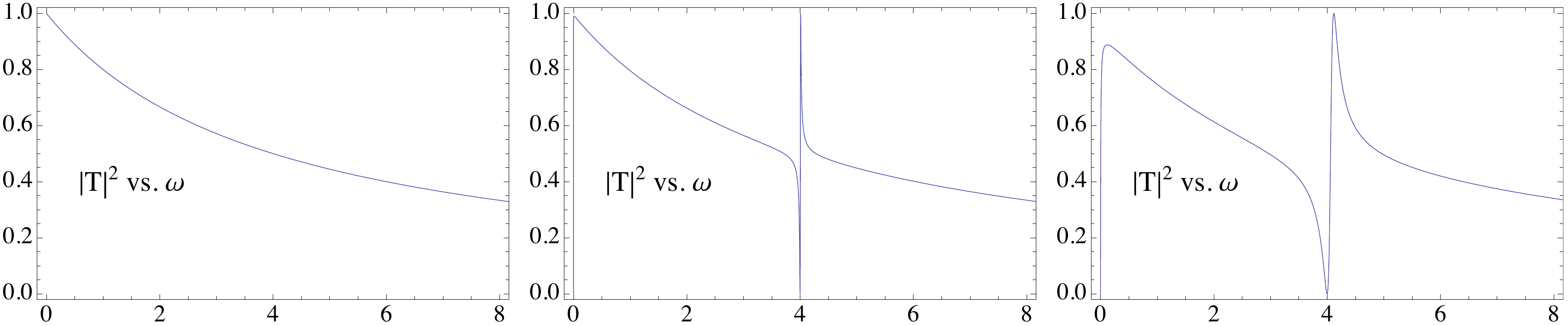}}
}
\caption{\small Transmission versus frequency for the linear system ($\lambda=0$) and normalized incident amplitude $J=1$ for three values of the coupling parameter $\gamma$: From left to right, $\gamma=0.0,0.2,0.7$.  The characteristic frequency of the oscillator is $\omega=E_0=4$.}
\label{fig:linear}
\end{figure}

The interaction between a resonant scatterer and an extended system that admits a spectral continuum of states is a fundamental problem in classical and quantum systems.  A variety of simple models have been devised to elucidate this interaction, and they have the advantage of providing clean mathematical treatments of specific phenomena.
H.~Lamb~\cite{Lamb1900} observed in 1900 that, if a simple harmonic oscillator is attached to a point on a string, the loss of energy by radiation into the string effectively results in the usual damped oscillator.   
  Komech~\cite{Komech1995} extended the analysis to a general simple nonlinear oscillator and proved that finite-energy solutions tend to an equilibrium state of the oscillator and in fact that transitions between any two equilibrium states are possible.
In the frequency domain, resonance effects of an oscillator on extended time-harmonic states were treated by Fano~\cite{Fano1961} (1961), in order  to explain peak-dip anomalies (the ``Fano line shape") observed in the scattering of electrons by the noble gases. These sharp resonances are a result of a weak coupling of a bound state to a continuum of extended states.  The frequency of the bound state is realized as an eigenvalue of the equations of the extended system, embedded in the continuous spectrum, and this eigenvalue is unstable with respect to perturbations of the system.  An analogous phenomenon occurs in the scattering of an EM plane wave by a photonic crystal slab.  Sharp peak-dip anomalies in the transmission of energy across the scatterer (similar to those in Fig.~\ref{fig:linear} and often referred to as resonances of Fano type) are due to the resonant excitation, by the incident wave, of the state that is localized in the slab \cite{ChristTikhodeevGippius2003,ChristZentgrafKuhl2004,ShipmanVenakides2003}.  These resonances can be analyzed rigorously by means of a complex-analytic perturbation theory of the scattering problem about the parameters of the bound state, and one obtains asymptotic formulae that reveal fine details of the anomalies~\cite{Shipman2010,ShipmanVenakides2005}.  The analysis is applicable quite generally to scattering problems that admit unstable bound states, including continuous as well as lattice models~\cite{ShipmanRibbeckSmith2010,PtitsynaShipman2012}. 

Kerr (cubic) nonlinearity in models of resonant harmonic scattering have been investigated by several authors, mostly numerically.  
Its effects on resonance in dielectric slabs is important for applications exploiting tunable bistability \cite{LousseVigneron2004}.
The Fano-Anderson model of a resonator coupled to a single chain of ``atoms" with nearest-neighbor interactions exhibits a sharp dip in the transmission coefficient in the middle of the spectrum.  When Kerr nonlinearity is introduced into the resonator, a stable scattering state bifurcates as the nonlinearity reaches a critical value, producing multiple scattering solutions and bistability in an interval of resonant frequencies~\cite{MiroshnicMingaleevFlach2005}.  Discrete nonlinear chains have been proposed for modeling photonic crystal waveguides \cite{McGurn2008,McGurnBirkok2004}.  When, instead of a single atomic chain, two atomic chains are coupled together, one can construct embedded trapped modes exhibiting the Fano-type anomaly observed in photonic systems, and numerical computations show that Kerr nonlinearity causes intricate multi-valued transmission coefficients~\cite{ShipmanRibbeckSmith2010}.

\begin{figure} 
\centerline{
    \scalebox{0.4}{\includegraphics{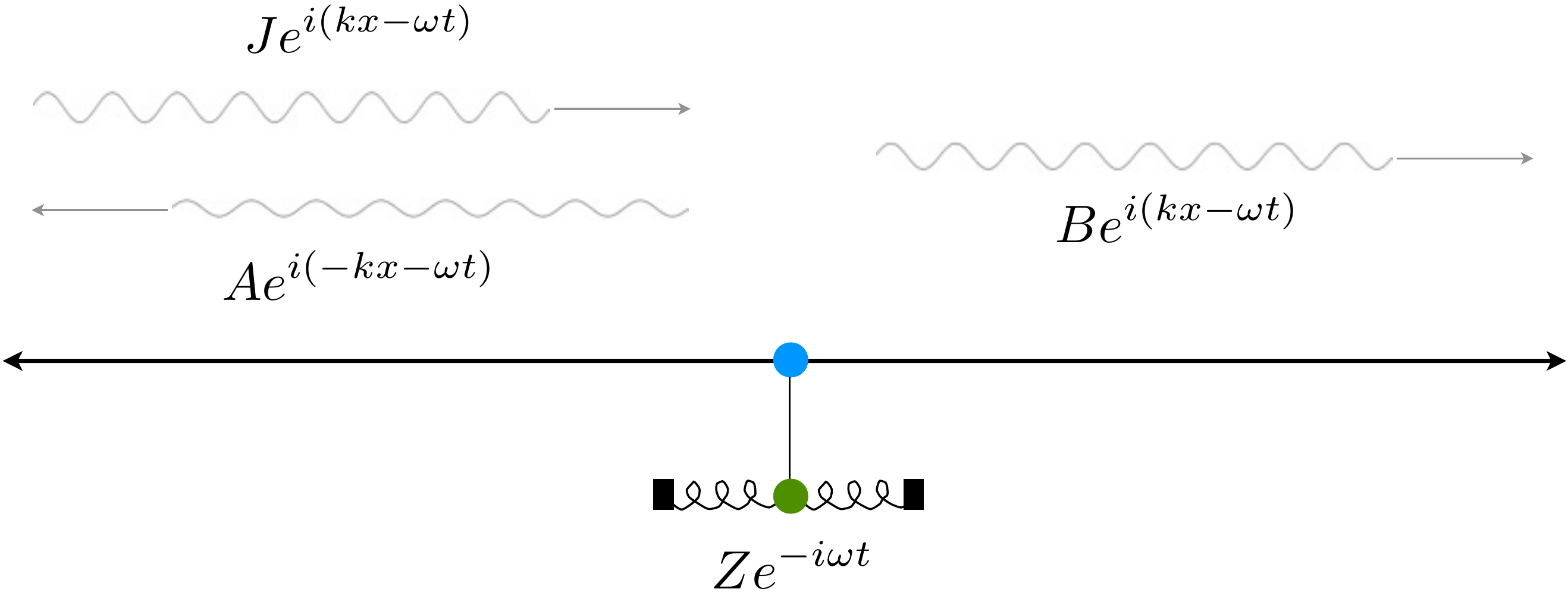}}
}
  \caption{\small The model for nonlinear resonant scattering.  The transmission line models the ambient space, the mass attached to the line models the (non-resonant) scatterer, and the nonlinear oscillator attached to the mass models the bound state, which is responsible for resonance.  The Kerr nonlinearity is confined to the oscillator.  The elements of a harmonic scattering field at frequency $\omega$ correspond to the form \eqref{scatteringsolution}.}
  \label{fig:model}
\end{figure}

In order to understand the fundamental effects that nonlinearity produces upon resonant scattering through exact explicit formulae, we introduce a dynamical system consisting of two systems coupled together (Fig.~\ref{fig:model}): (1) a transmission line of Schr\" odinger type with a point-mass defect acting as a non-resonant scatterer and (2) a nonlinear resonator coupled to the defect on the line.  
When the nonlinearity is set to zero, the model incorporates the essential features of these linear photonic systems described above  inasmuch as the resonant phenomena are concerned.  The model elucidates fundamental nonlinear effects that coincide with asymptotic relations between the parameters of coupling and nonlinearity.
A discussion and comparison with the models of Lamb and Komech as well as the Duffing oscillator is offered in the final section~\ref{sec:discussion}.  The reader may decide to browse that discussion before embarking on the analysis of our model in sections \ref{sec:harmonic}--\ref{sec:stability}.

\smallskip

The transmission line with a point mass at $x=0$ is the system
\begin{equation*}
  \renewcommand{\arraystretch}{1.2}
\left.
  \begin{array}{l}
    iu_t(x,t) = -hu_{xx}(x,t), \quad x\not=0,\\
       iu_t(0,t) = - \tau\left(u_x(0^+,t)-u_x(0^-,t)\right),
  \end{array}
\right.
\end{equation*}
in which $h>0$ carries units of area per time and $\tau>0$ carries units of length per time.  Selecting $T=h/\tau^2$ and $X=h/\tau$ as units of time and space nondimensionalizes the equation and removes $h$ and $\tau$.  Thus the nondimensional frequency $\omega=1$ corresponds to the dimensional frequency $2\pi\tau^2/h$.
The resonator is a nonlinear harmonic oscillator with cubic, or Kerr, nonlinearity, that obeys the equation in nondimensional form
\begin{equation*}
  i\dot{z}(t) = E_0 z(t) + \lambda|z(t)|^2 z(t).
\end{equation*}
The nondimensional characteristic frequency $E_0$ is relative to the fixed frequency unit $2\pi/T$, and $\lambda$ is a nondimensionalization of a parameter carrying units of frequency per square units of $z$ (the values $u$ and $z$ of the fields are assumed to be nondimensionalized relative to a common unit).
Neither $E_0$ nor $\lambda$ can be eliminated from the equation.

These two systems are then coupled by a parameter $\gamma$,
\begin{equation}\label{system}
  \renewcommand{\arraystretch}{1.2}
\left.
  \begin{array}{ll}
    iu_t(x,t) = -u_{xx}(x,t), \ \ \ \ \ \ x\ne 0, & \text{(transmission line)}\\
       i\dot y(t) = \gamma z(t) - (u_x(0^+,t)-u_x(0^-,t)), \quad y(t)=u(0,t)
     & \text{(point defect on the line)}\\    
    i\dot{z}(t) = E_0 z(t) + \gamma\,y(t) + \lambda|z(t)|^2 z(t). & \text{(resonator)}
  \end{array}
\right.
\end{equation}
Nonlinearity is measured by the composite parameter
\begin{equation*}
  \mu=\lambda J^2
\end{equation*}
where $J$ is the amplitude of a monochromatic field emanating from a source at $-\infty$ (see Fig.~\ref{fig:model} or \eqref{scatteringsolution}).  In its dimensional form, $\mu$ and $\gamma$ are frequencies, with $\mu$ being amplitude-dependent and depending quadratically on the strength of the incident field.  

The  resonant amplification of the oscillator in the linear system ($\lambda=0$), accompanied by sharp transmission anomalies near the resonant frequency $E_0$, as described above, is portrayed in Fig.~\ref{fig:linear}.  As the coupling $\gamma$ of the oscillator to the transmission line vanishes, the resonant amplification becomes unbounded.  In the limit $\gamma\to0$, the motion of the oscillator becomes a bound state for the decoupled system, whose frequency is embedded in the continuous spectrum arising from the transmission line.

The nonlinear system \eqref{system} admits time-periodic solutions, which the Kerr form $\lambda|z|^2z$ of the nonlinearity allows to be  monochromatic. 
Even a small  nonlinearity has a pronounced effect on the resonance, due to the high-amplitude fields produced.  Fig.~\ref{fig:transmission} shows how multiple scattering solutions (nonuniqueness of the scattering problem) appear near the resonant frequency and spread to higher frequencies as the nonlinearity parameter is increased. Similarly, a new branch of solutions emerges from the infinite frequency limit and spreads to lower frequencies.  

The present work proves that the bifurcations occur as portrayed in Fig.~\ref{fig:bifurcation}, as the nonlinearity increases from the value zero.  The analysis reveals characteristic power laws between the coupling parameter $\gamma$ and the composite parameter of nonlinearity $\mu$ in the asymptotic regime of $\gamma\to0$ and $\mu\to0$ (see Table~\ref{table:bifurcation} and Fig.~\ref{fig:bifurcation}).  The critical relation $\gamma^4/\mu\sim\text{\itshape Const.}$ marks the onset of a narrow frequency interval $[\omega_1,\omega_2]$ of triple solutions near the resonant frequency.  For each frequency above this interval and below a high frequency $\omega_3\sim C\gamma^{-2}$, the system admits a unique harmonic solution.  As $\mu$ increases relative to $\gamma$, the intervals $[\omega_1,\omega_2]$ and $[\omega_3,\infty)$ of triple solutions widen until they merge, that is $\omega_2=\omega_3\sim C$, at the critical power law $\gamma^2/\mu\sim\text{\itshape Const\,}$.
For any fixed, small $\mu$ and $\gamma$, there are at most three frequencies that separate intervals of unicity from intervals of triple solutions; we call these {\itshape transition frequencies}. 
Understanding of the dependence of the transition frequencies $\omega_1$, $\omega_2$, and $\omega_3$ on the parameters $\gamma$ and $\mu$ can be illustrated graphically through the intersection points of two curves in the plane whose coordinates are reparameterizations of the frequency of a harmonic excitation and the ``response" of the resonator (\S\ref{sec:bifurcation}).

The stability of multiple scattering solutions is important in applications involving bistable optical transmission \cite{CowanYoung2003,LousseVigneron2004,MiroshnicMingaleevFlach2005,YanikFanSoljacic2003}.
It is commonly understood that, at frequencies for which a nonlinear resonant system admits a unique periodic solution (for a given system and source-field amplitude $J$), this field is stable under perturbations.  At frequencies for which there are exactly three periodic solutions, there should be bistability, that is, two solutions are stable and one unstable, the unstable one being that with intermediate amplitude (see, {\itshape e.g.}, the references just cited).
The analysis in this section yields precise statements about this kind of bistable behavior in our model of nonlinear resonance (Theorem~\ref{thm:stability}).

\medskip

\section{Harmonic solutions of the nonlinear model}\label{sec:harmonic}  

System (\ref{system}) admits harmonic scattering solutions, in which a field $Je^{i(kx-\omega t)}$ oscillating at frequency $\omega$ in the string is incident on the resonator from the left (Fig.~\ref{fig:model}),
\begin{equation}\label{scatteringsolution}
  \renewcommand{\arraystretch}{1.2}
\left.
  \begin{array}{ll}
    u(x,t) = (Je^{ikx} + Ae^{-ikx})e^{-i\omega t}, & x<0, \\
    u(x,t) = Be^{ikx}e^{-i\omega t}, & x>0, \\
    y(t) = Be^{-i\omega t}, \\
    z(t) = Ze^{-i\omega t}.
  \end{array}
\right.
\end{equation}
where $y(t)=u(0,t)$ represents the state of the defect in the transmission line. 
The wave number $k>0$ in the string and the frequency $\omega$ are subject to the dispersion relation for the free string $\omega = k^2$.
The continuity of $u$ provides the relation
\begin{equation*}
  J + A = B,
\end{equation*}
and \eqref{system} reduces to the algebraic system
\begin{eqnarray*}
  && (2ik - k^2)B + \gamma Z = 2ikJ\,, \\
  && \gamma B + (E_0 - k^2)Z + \lambda |Z|^2 Z = 0\,,
\end{eqnarray*}
which can be written as a cubic equation for $Z/J$ and linear a relation between $B/J$ and $Z/J$\,:
\begin{equation}\label{algebraicsystem}
  \renewcommand{\arraystretch}{1.5}
\left.
  \begin{array}{l}
    \displaystyle \frac{2ik\gamma}{2ik-k^2}
    - \left( \frac{\gamma^2}{2ik-k^2} + k^2-E_0 \right)\frac{Z}{J}
    + \lambda J^2 \left|\frac{Z}{J}\right|^2 \frac{Z}{J} \ =\ 0\,, \\
    \vspace{-2ex}\\
    \displaystyle \frac{B}{J} \ =\ \frac{2ik-\gamma (Z/J)}{2ik - k^2}\,.
  \end{array}
\right.
\end{equation}
Evidently the dependence on $\lambda$ and $J$ is only through the composite parameter
\begin{equation*}
  \mu := \lambda J^2.
\end{equation*}

\begin{figure}
\begin{center}
\scalebox{0.35}{\includegraphics{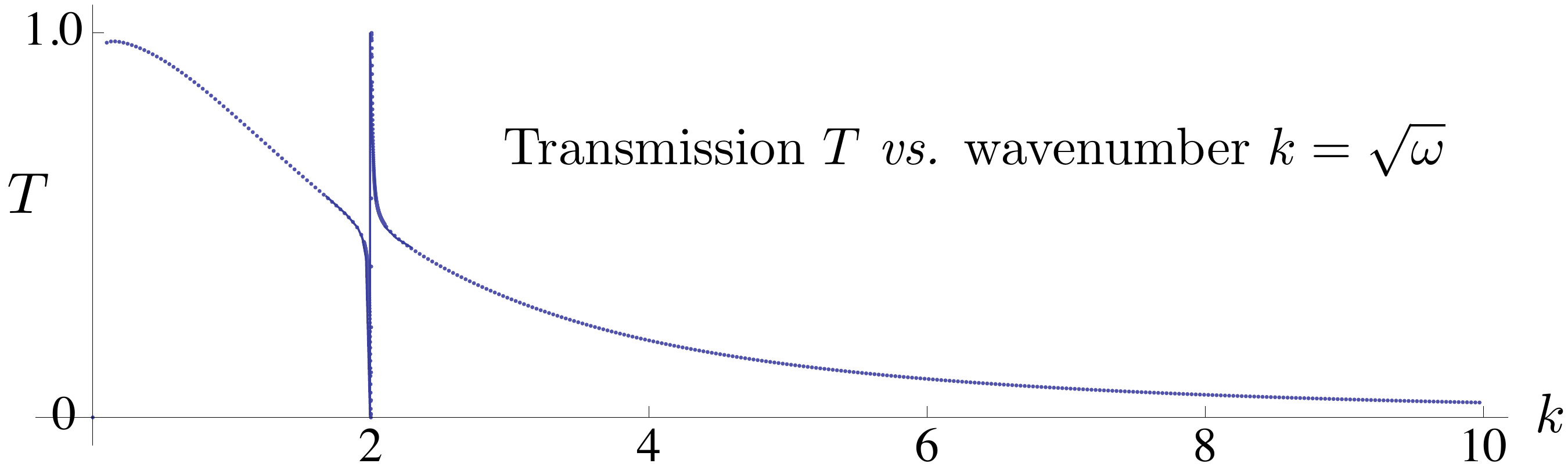}}\\
\scalebox{0.35}{\includegraphics{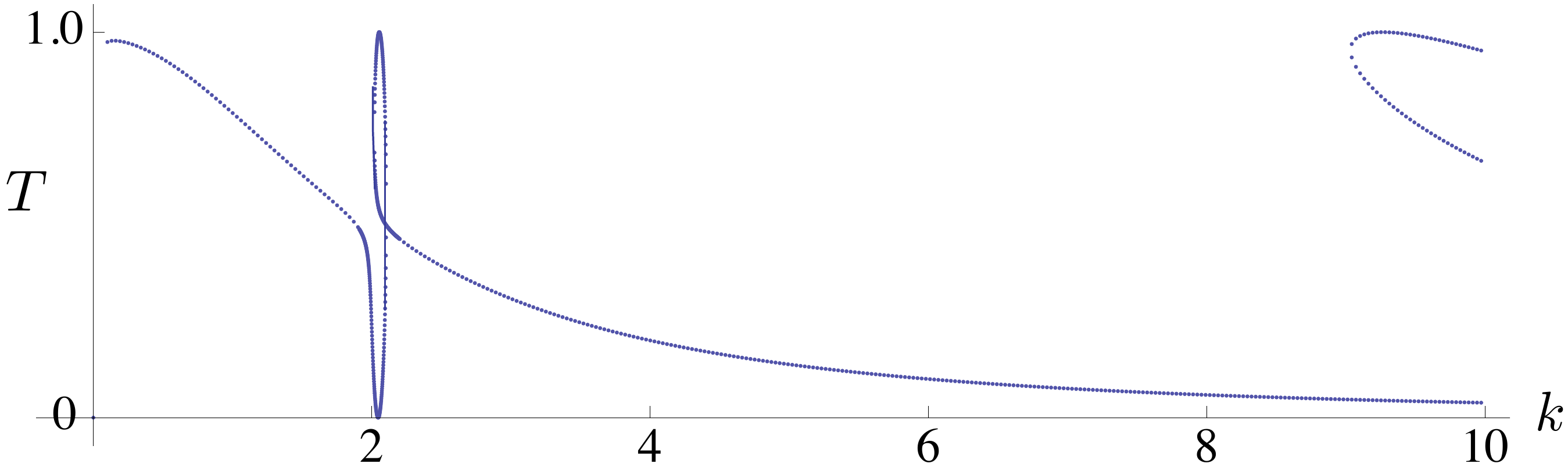}}\\
\scalebox{0.35}{\includegraphics{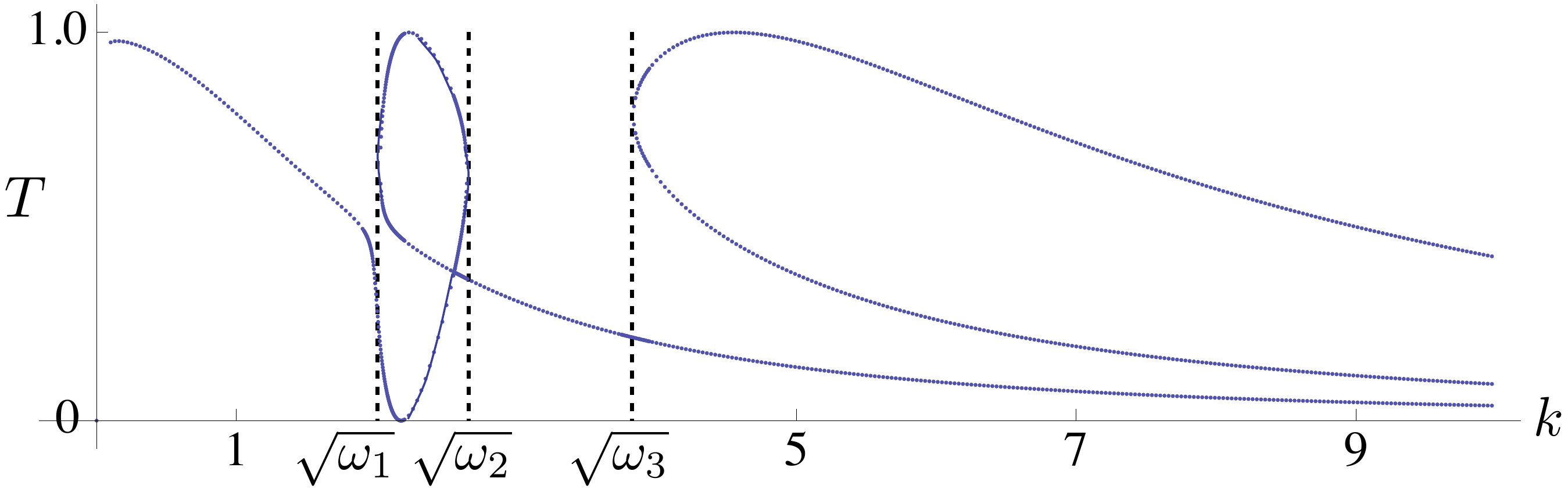}}\\
\scalebox{0.35}{\includegraphics{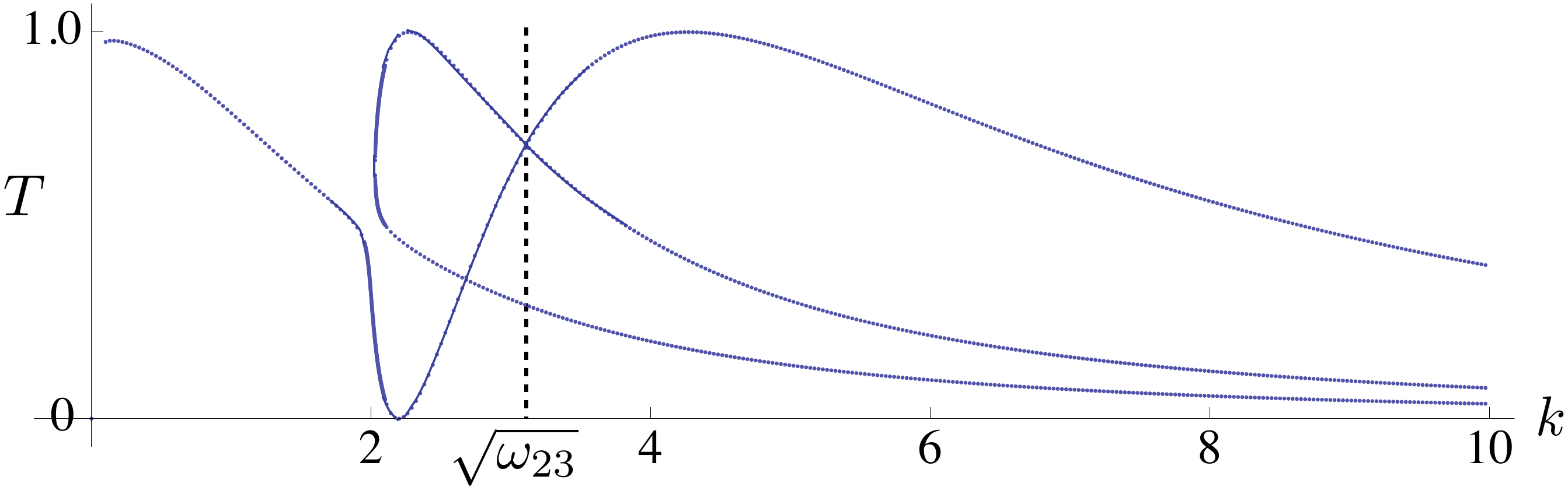}}\\
\scalebox{0.35}{\includegraphics{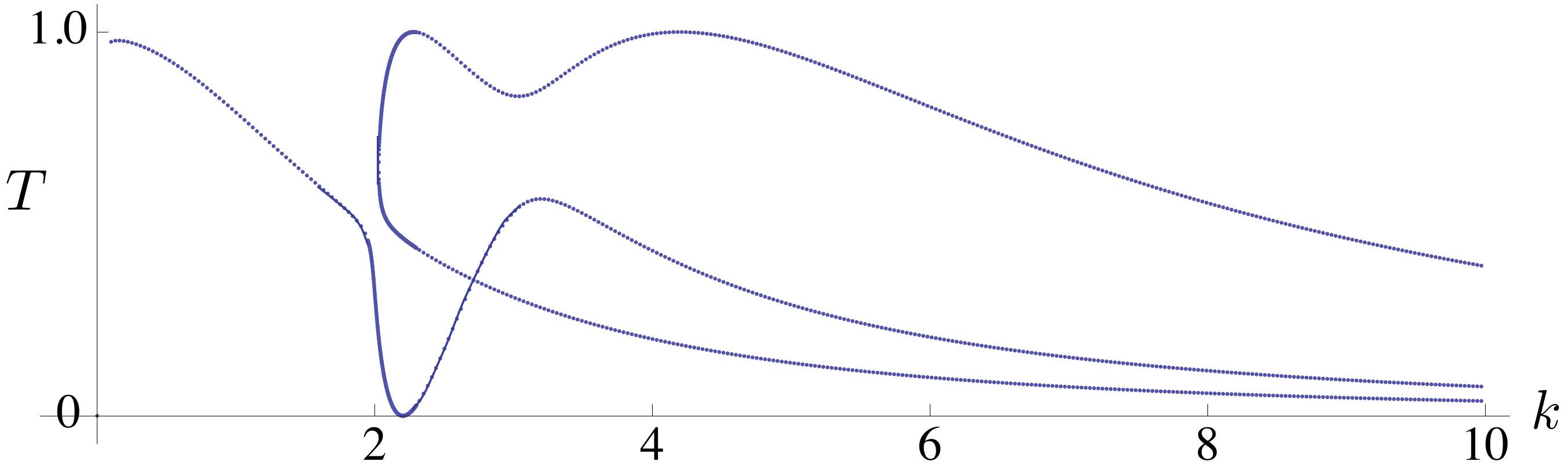}}
\end{center}
\caption{\small
The transmission coefficient $T:=|B/J|^2$ of the harmonic scattering solutions \eqref{scatteringsolution} of the string-oscillator system \eqref{system} {\itshape vs.} the wavenumber $k=\sqrt{\omega\,}\,$.  The graphs show progression of values of the composite nonlinearity parameter $\mu=\lambda J^2$ (from top to bottom, $\mu = 0,\, 0.001,\, 0.0035,\, 0.003856,\, 0.00395$), all other parameters being fixed ($E_0=4$, $\gamma=0.3$).
The top graph shows the linear case, in which each frequency admits a single harmonic solution.  The sharp anomaly occurs near the characteristic frequency $E_0=4$ of the oscillator.
For nonzero $\mu$, there is a half-infinite frequency interval $(\omega_3,\infty)$ of triple solutions.
As $\mu$ increases (or as $\gamma$ decreases), a narrow frequency interval $(\omega_1,\omega_2)$ of triple solutions emerges; this is the $\omega_1$-$\omega_2$ bifurcation.  Meanwhile, $\omega_3$ decreases, eventually merging with $\omega_2$ and eliminating the interval $(\omega_2,\omega_3)$ of unique solutions; this is the $\omega_2$-$\omega_3$ bifurcation.  A diagram of the bifurcations for this progression is shown in Fig.~\ref{fig:bifurcation} (bottom).
}
\label{fig:transmission}
\end{figure}

\begin{figure}
\begin{center}
\scalebox{0.32}{\includegraphics{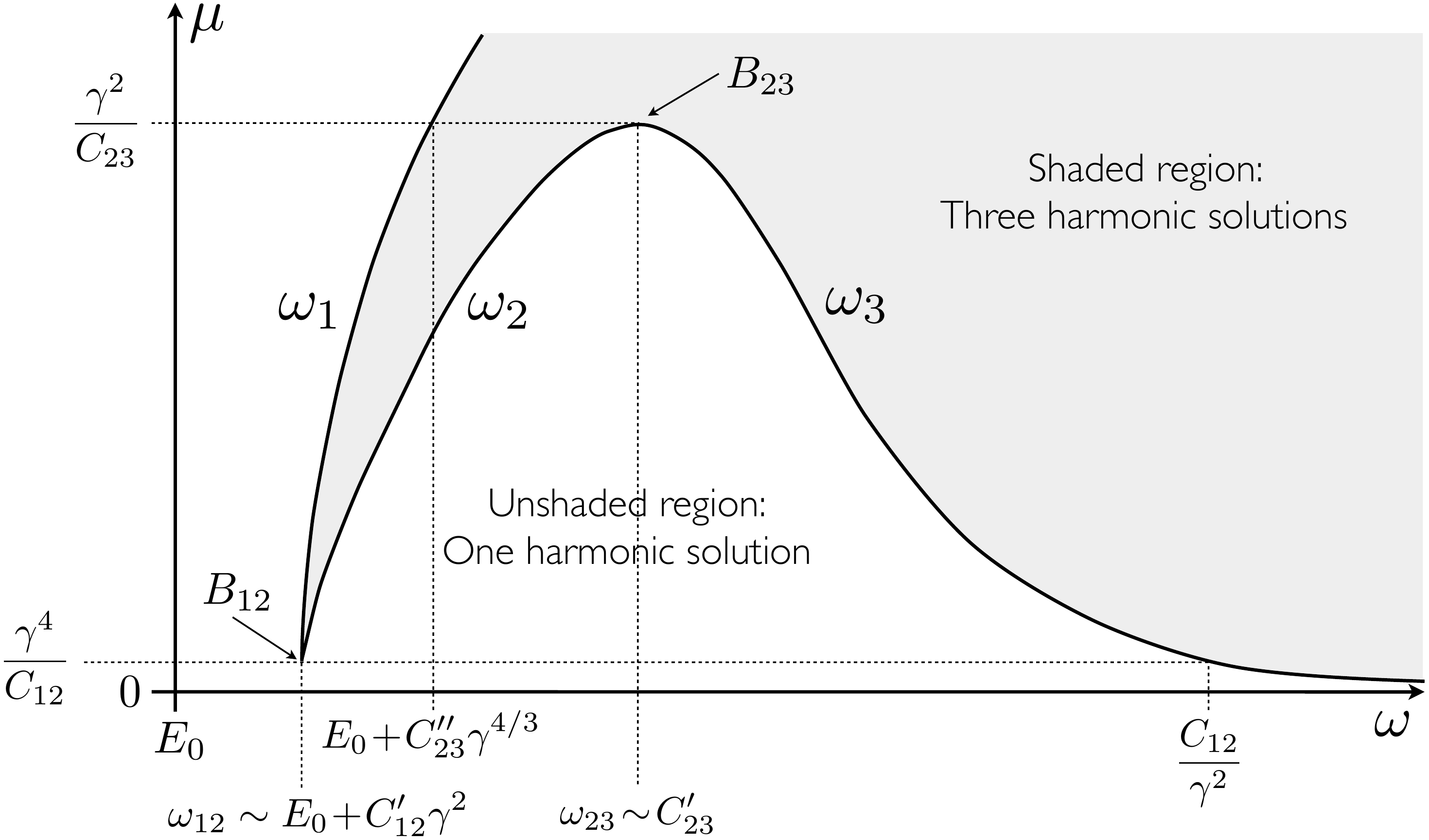}}\\
\vspace{2ex}
\scalebox{0.35}{\includegraphics{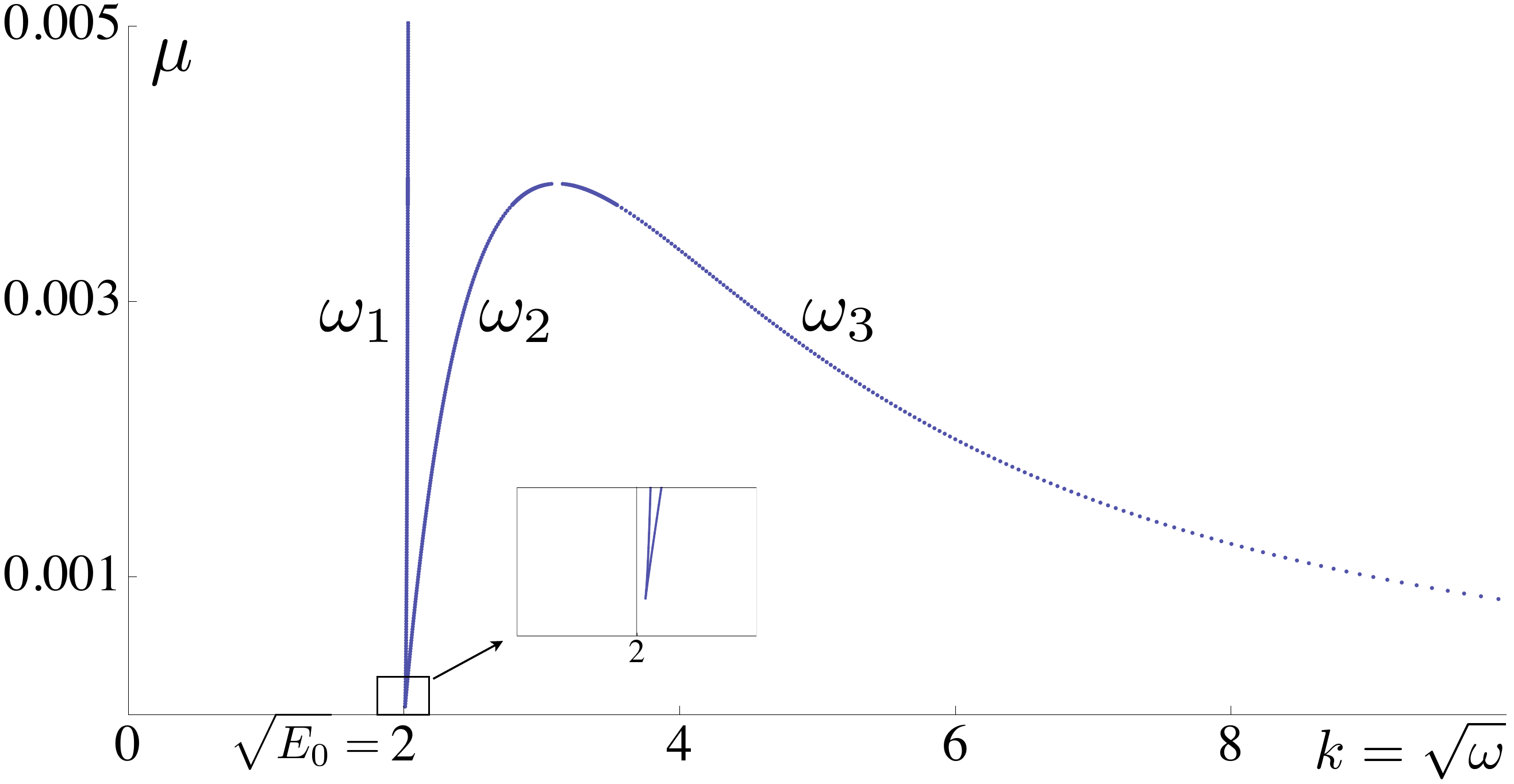}}\\
\end{center}
\caption{\small {\bfseries Top:} Transition frequencies $\omega_1$, $\omega_2$, and $\omega_3$, {\em vs.}~$\mu$, at which the number of harmonic scattering solutions of the form \eqref{scatteringsolution} changes from $1$ to $3$ or vice-versa, according to whether the cubic polynomial \eqref{Polynomial} for the response has one or three real roots.
The unshaded region indicates those $(\omega,\mu)$ pairs that admit only one real root, whereas pairs in the lightly shaded region admit three real roots.
Bifurcations occur at the points $B_{12}$ (the cuspidal $\omega_1$-$\omega_2$ bifurcation in the text) and $B_{23}$ ($\omega_2$-$\omega_3$ bifurcation) as $\mu$ varies.
The coordinates of the two bifurcations are labeled with their asymptotic values for small $\gamma$ and $\mu$, according to Table \ref{table:bifurcation}.
{\bfseries Bottom:} 
This is a numerical computation for $E_0=4$ and $\gamma=0.3$ that corresponds to Fig.~\ref{fig:transmission}.
}
\label{fig:bifurcation}
\end{figure}

\begin{table}  
\centerline{
\begin{tabular}{|c|c|l|}
bifurcation
& power law
& \parbox{8em}{\centerline{$\omega_1$}}\parbox{0pt}{$\big|$}\parbox{8em}{\centerline{$\omega_2$}}\parbox{0pt}{$\big|$}\parbox{8em}{\centerline{$\omega_3$}} \\
\hline \hline
$\omega_1=\omega_2$
& $\displaystyle\frac{\gamma^4}{\mu}\sim C_{12}$
& \parbox{16em}{\centerline{$\omega_{12}-E_0 \,\sim\, C'_{12}\,\gamma^2$}}\parbox{0pt}{$\Bigg|$}\parbox{8em}{\centerline{$\displaystyle\omega_3\,\sim\, \frac{C_{12}}{\gamma^2}$}} \\
\hline
$\omega_2=\omega_3$
& $\displaystyle\frac{\gamma^2}{\mu}\sim C_{23}$
& \parbox{9em}{\centerline{$\omega_1-E_0\,\sim\, C''_{23}\,\gamma^{4/3}$}}\parbox{0pt}{$\Bigg|$}\parbox{16em}{\centerline{$\omega_{23}\,\sim\,C'_{23}$}} \\
\hline
\end{tabular}}
\caption{\small  Asymptotics of the bifurcations of the transition frequencies for small $\gamma$ and $\mu$.  The bifurcation diagram with $\mu$ as the bifurcation parameter and fixed $\gamma$ is shown in Fig.~\ref{fig:bifurcation}.
All constants depend only on $E_0$ and are computed explicitly in the text.  The three relations in the first row are given in equations (\ref{12bifurcation},\ref{omega12},\ref{omega3at12}); the relations in the second row are given in equations (\ref{23bifurcation},\ref{omega1at23},\ref{omega23}).}
\label{table:bifurcation}
\end{table}

\begin{figure}
\begin{center}
\scalebox{0.5}{\includegraphics{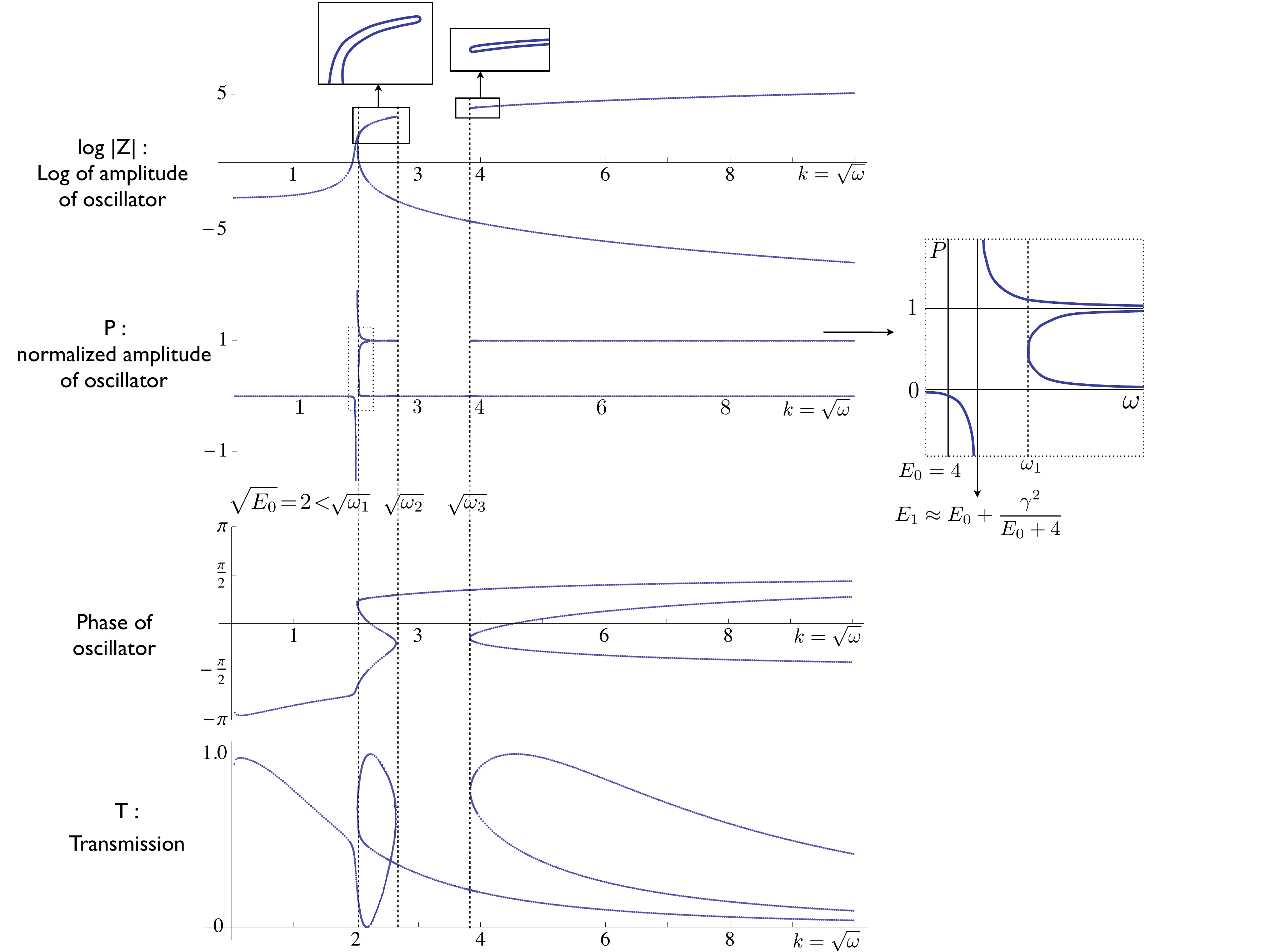}}
\vspace{-2ex}
\end{center}
\caption{\small 
These graphs show the harmonic scattering states \eqref{scatteringsolution} of the string-oscillator system \eqref{system} {\em vs.} the wavenumber $k\!=\!\sqrt{\omega}$.
Plotted from top to bottom are
{\bfseries (i)} the logarithm $\log|Z|$ of the amplitude of the oscillator's motion $z(t)=Ze^{-i\omega t}$;
{\bfseries (ii)} a rescaling $P$ of the response ${\R}=|Z/J|^2$ defined by~\eqref{RP} ($J$~is the amplitude of the incident field) with detail near the resonant frequency depicted to the right;
{\bfseries (iii)} the phase $\arg(Z)$ of the oscillator; and
{\bfseries (iv)} the transmission coefficient $T\!=\!|B/J|^2$.
For the parameter values $E_0\!=\!4.0$, $\gamma\!=\!0.3$, and $\mu\!=\!0.0035$ chosen here,
there are two frequency intervals $(\omega_1,\omega_2)$ and $(\omega_3,\infty)$, delineated by the vertical dotted lines, in which the system has three solutions; at frequencies in the complementary intervals $(0,\omega_1)$ and $(\omega_2,\omega_3)$, the system has only one solution.
The two high-amplitude oscillations have almost identical amplitudes, as shown in the top graph, but they differ by a phase that tends to $\pi$ as $\omega\to\infty$, as seen in the third graph.
A graph of the transition frequencies {\em vs.}~$\mu$ is shown in Fig.~\ref{fig:bifurcation}.
}
\label{fig:ZTPgraphs}
\end{figure}

In Figures.~\ref{fig:transmission} and \ref{fig:ZTPgraphs}, one observes two intervals of the $\omega$-line in which the harmonic scattering problem has three solutions and two intervals in which it has one solution.  These intervals are separated by three transition frequencies $k_i^2=\omega_i(\gamma,\mu,E_0)$, $i=1,2,3$\,:
\begin{equation}\label{}
  \renewcommand{\arraystretch}{1.2}
\left.
  \begin{array}{ll}
    (0,\omega_1)\cup(\omega_2,\omega_3) & \text{one solution,} \\
    (\omega_1,\omega_2)\cup(\omega_3,\infty) & \text{three solutions.}
  \end{array}
\right.
\end{equation}
When $\mu$ is decreased or $\gamma$ increased, the points $\omega_1$ and $\omega_2$ approach each other and are annihilated, and there remains a single interval of one solution and one of three solutions, separated by the point $\omega_3$.  We will call this threshold the $\omega_1$-$\omega_2$ bifurcation.  On the other hand, increasing $\mu$ or decreasing $\gamma$ brings the points $\omega_2$ and $\omega_3$ together, resulting in a single interval $[\omega_1,\infty)$ of multiple solutions.  We call the threshold that occurs when they are annihilated the $\omega_2$-$\omega_3$ bifurcation.

\medskip

Each of these bifurcations is analyzed in detail in the following section, and Table~\ref{table:bifurcation} is a summary of the results.  Asymptotically, as $\mu$ and $\gamma$ vanish, the bifurcations are characterized by specific power laws.  The $\omega_1$-$\omega_2$ bifurcation occurs at relatively weak nonlinearity, $\mu\sim C\gamma^4$, when its effects on the resonance near $E_0$ first appear.  The $\omega_2$-$\omega_3$ bifurcation occurs at much higher values of nonlinearity relative to the coupling parameter.  These constants as well as those in the asymptotics for the frequencies $\omega_i$ are computed explicitly in terms of the resonant frequency $E_0$ alone.

Analysis of the number of scattering solutions as a function of frequency is facilitated by a polynomial equation for the square of the amplitude enhancement $\R=|Z/J|^2$ in the resonator, which we call the {\em response}.  The first equation of \eqref{algebraicsystem} yields
\begin{equation}\label{Requation}
  \R\left(\mu \R + \frac{\gamma^2}{\omega+4} - (\omega-E_0) \right)^2
  + \frac{4\gamma^4}{\omega(\omega+4)^2}\R - \frac{4\gamma^2}{\omega+4} \,=\, 0.
\end{equation}
The roots $\R$ of this equation, which are necessarily positive, are in one-to-one correspondence with the solutions $(B,Z)$ of \eqref{algebraicsystem}.  This equation can have multiple roots only if
\begin{equation}\label{omegainequality}
   (\omega-E_0)(\omega+4) - \gamma^2 > 0, \qquad \text{(necessary for multiple solutions)}
\end{equation}
in particular, if $\omega>E_0$, that is, if $\omega$ exceeds the frequency of the free resonator.
The change of variable
\begin{equation}\label{RP}
  \R = \frac{P}{\mu}\left( \omega-E_0 - \frac{\gamma^2}{\omega+4} \right)
\end{equation}
transforms equation \eqref{Requation} into
\begin{equation}\label{Polynomial}
  \boxed{f(P):= P(P-1)^2 + \alpha P - \beta = 0\,,}
\end{equation}
in which
\begin{equation}\label{alphabeta}
  \renewcommand{\arraystretch}{1.5}
\left.
  \begin{array}{l}
    \displaystyle \alpha =\alpha(\omega;\gamma,E_0) := \frac{1}{\omega}\left( \frac{2\gamma^2}{(\omega-E_0)(\omega+4)-\gamma^2} \right)^2,\\
    \vspace{-2ex}\\
    \displaystyle \beta =\beta(\omega;\gamma,\mu,E_0) := \frac{4\gamma^2\mu(\omega+4)^2}{((\omega-E_0)(\omega+4)-\gamma^2)^3}\,.
  \end{array}
\right.
\end{equation}
The quantities $\R$, $\alpha$, and $\beta$ can be written more compactly using the notation
\begin{equation*}
  \renewcommand{\arraystretch}{1}
\left.
  \begin{array}{ll}
    \rho := \omega+4 & \sim \omega\, \quad \text{as } \omega\to\infty, \\
    \sigma := \omega-E_0 - \gamma^2/\rho & \sim \omega\, \quad \text{as } \omega\to\infty.
  \end{array}
\right.
\end{equation*}
%
%
\begin{equation*}
  \R = \frac{\sigma}{\mu}\, P\,, \quad
  \alpha = \frac{4\gamma^4}{\omega\sigma^2\rho^2}, \quad
  \beta = \frac{4\gamma^2\mu}{\sigma^3\rho}.
\end{equation*}

If one fixes the parameters $\mu$, $\gamma$, and $E_0$ of the dynamical system, one can consider the roots of the polynomial $f(P)$ as functions of frequency $\omega$.
These roots correspond to all possible responses $\R$ associated with harmonic scattering solutions at $\omega$.  Transition frequencies separating unicity of the response from multiple responses occur when $f(P)$ has a double root.
Thus {\em analysis of transition frequencies is tantamount to analysis of the parameters for which $f(P)$ has a double root}.

\section{Multiple harmonic solutions and their bifurcations}\label{sec:bifurcation} 
 
We have seen that certain transition frequencies $\omega_i(\mu,\gamma;E_0)$ separate intervals of unique response to harmonic forcing by the source field $Je^{-i\omega t}$ from intervals of triple response.  The diagram in Fig.~\ref{fig:bifurcation} shows how the transition frequencies merge or separate as a function of the nonlinearity in the asymptotic regime of $\gamma\to0$ and $\mu\to0$.  This section is devoted to proving the asymptotic power laws in that diagram.  The main results are stated in the following theorem.

\begin{Theorem}\label{thm:bifurcation}
The transition frequencies $\omega_i$ that separate frequency intervals of unique scattering solutions from intervals of multiple solutions vary with the nonlinearity parameter $\mu$ according to the diagram in Fig.~\ref{fig:bifurcation} if the coupling parameter $\gamma$ is sufficiently small.
More precisely, \\

\vspace{-2ex}
\noindent
{\bfseries 1.}\; There are at most three transition frequencies if $\gamma$ is sufficiently small.\\
{\bfseries 2.}\; There is a point $B_{12}=(\omega_{12},\mu_*)$ and positive numbers $C_{12}$ and $C'_{12}$ with
\begin{equation*}
 \frac{\gamma^4}{\mu_*}\sim C_{12} \quad\text{and}\quad \omega_{12}-E_0\sim C'_{12}\gamma^2 \quad (\gamma\to0)
\end{equation*}
such that, if $\mu<\mu_*$, there is a single transition frequency $\omega_3$ and if $\mu>\mu_*$ but $\mu$ is not too large, there are three transition frequencies with $\omega_1<\omega_2<\omega_3$.  As $\mu\to\mu_*$ from above, $\omega_1\to\omega_{12}$ and $\omega_2\to\omega_{12}$.  Moreover when $\mu=\mu_*$, there are two transition frequencies $\omega_{12}=\omega_1=\omega_2<\omega_3$ and
\begin{equation*}
  \omega_3\sim \frac{C_{12}}{\gamma^2} \quad (\gamma\to0).  \qquad (\omega_1=\omega_2)
\end{equation*}
The numbers $C_{12}$ and $C'_{12}$ depend only on $E_0$ and are given by \eqref{12bifurcation} and \eqref{omega12}. \\
{\bfseries 3.}\; There is a point $B_{23}=(\omega_{23},\mu^*)$ and positive numbers $C_{23}$ and $C'_{23}$ with
\begin{equation*}
  \frac{\gamma^2}{\mu^*}\sim C_{23} \quad\text{and}\quad \omega_{23}\sim C'_{23}
   \quad (\gamma\to0)
\end{equation*}
such that, if $\mu>\mu^*$, there is a single transition frequency $\omega_1$ and
if $\mu<\mu^*$ but $\mu$ is not too small, there are three transition frequencies with $\omega_1<\omega_2<\omega_3$.  As $\mu\to\mu^*$ from below, $\omega_2\to\omega_{23}$ and $\omega_3\to\omega_{23}$.  Moreover, when $\mu=\mu^*$, there are two transition frequencies $
\omega_1<\omega_2=\omega_3=\omega_{23}$ and
\begin{equation*}
  \omega_1-E_0 \sim \frac{3}{C_{23}^{1/3}}\,\gamma^{4/3}  \quad (\gamma\to0).
\end{equation*}
The numbers $C_{23}$ and $C'_{23}$ depend only on $E_0$ and are given by \eqref{23bifurcation} and \eqref{omega23}. \\
\end{Theorem}

The transition frequencies are characterized by 
 the property that there exists a real number $P$ such that both \eqref{Polynomial} and its derivative with respect to $P$ vanish.  The system of the two conditions is algebraically equivalent to the pair
\begin{equation}\label{Pomegasystem}
\renewcommand{\arraystretch}{1}
\boxed{\left.
  \begin{array}{rcl}
    \alpha(\omega;\gamma,E_0) &=& -3(P-\frac{1}{3})(P-1)\,,\\
    \vspace{-1.5ex}\\
    \beta(\omega;\gamma,\mu,E_0) &=& -2P^2(P-1)\,.
  \end{array}
\right.}
\quad \text{(conditions for transition frequencies)}
\end{equation}
%

It is convenient to work with the roots of $\,\sigma\rho = (\omega-E_0)(\omega+4)-\gamma^2 = (\omega-E_1)(\omega+c)$, which are small perturbations of $E_0$ and $-4$ as $\gamma\to0$,
\begin{eqnarray}
  (\omega-E_1)(\omega+c) &=& (\omega-E_0)(\omega+4)-\gamma^2, \\
  E_1 &=& E_0+\epsilon\,,\\
  c &=& 4 + \epsilon\,, \\
  \epsilon &=& \frac{\gamma^2}{E_0+4} + {\cal O}(\gamma^4)\,, \\
  a = 4+E_1 &=& 4+E_0+\epsilon\,.
\end{eqnarray}
Because of inequality (\ref{omegainequality}), multiple solutions are possible only for $\omega>E_1$, and we therefore introduce the variables
\begin{eqnarray}
  && \nu = \omega - E_1, \\
  && \tau = \nu^{-1}, \\
  && q = 3P-2\,.
\end{eqnarray}
The algebraic system \eqref{Pomegasystem} in $P$ and $\omega$ can be rewritten as a system in $q$ and $\tau$:
\begin{eqnarray*}
  && (1-q)(1+q) = 3\alpha = 12\gamma^4 \frac{\tau^5}{(1+E_1\tau)(1+(a+\epsilon)\tau)^2}, \\
  && (1-q)(2+q)^2 = \frac{27}{2} \beta = \frac{54\gamma^2\mu\,\tau^4(1+a\tau)^2}{(1+(a+\epsilon)\tau)^3}.
\end{eqnarray*}
Dividing the second by the first and retaining the first equation yields the equivalent pair
\begin{equation}\label{qtausystem}
\boxed{\renewcommand{\arraystretch}{1}
\left\{
  \begin{array}{rcll}
  (1-q)(1+q) &=& \displaystyle 12\gamma^4 \frac{\tau^5}{(1+E_1\tau)(1+(a+\epsilon)\tau)^2} 
              =  \frac{12\gamma^4}{\nu^2(\nu+E_1)(\nu+(a+\epsilon))^2}\,, & (a) \\
    \vspace{-1ex} \\
    \displaystyle \frac{(2+q)^2}{1+q} &=& \displaystyle \frac{9}{2}\frac{\mu}{\gamma^2}\frac{(1+a\tau)^2(1+E_1\tau)}{\tau(1+(a+\epsilon)\tau)}
              = \frac{9}{2}\frac{\mu}{\gamma^2}\frac{(\nu+a)^2(\nu+E_1)}{\nu(\nu+(a+\epsilon))}\,. & (b)
  \end{array}
\right.}
\end{equation}

\subsection{Graphical depiction of transition frequencies and their bifurcations}\label{subsec:graphical}

\begin{figure}
\centerline{
\scalebox{0.44}{\includegraphics{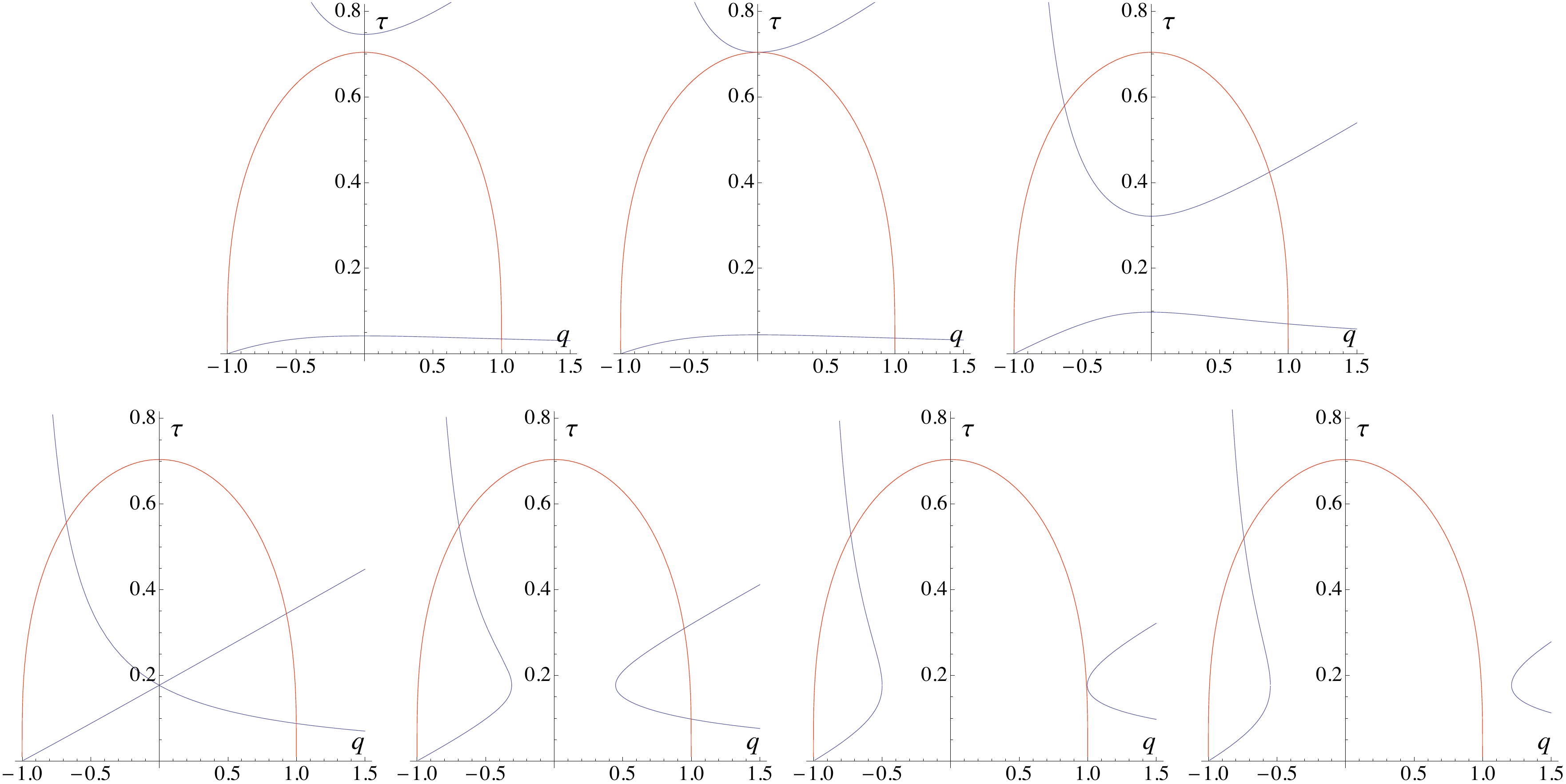}}
}
\caption{\small The relations (\ref{qtausystem}) in the $(q,\tau)$ plane, whose intersections give the transition frequencies through $\omega=E_1+\tau^{-1}$.  The symmetric one is (\ref{qtausystem}a).  The values of $\gamma$ and $E_0$ are fixed, and the graphs from left to right show the evolution of the relation (\ref{qtausystem}b) as $\mu$ increases.  The two bifurcations occur at the tangential intersections, as in the second ($\omega_1$-$\omega_2$ bifurcation) and fifth ($\omega_2$-$\omega_3$ bifurcation) graphs.}
\label{fig:Graph}
\end{figure}

\begin{figure} 
\centerline{
\scalebox{0.48}{\includegraphics{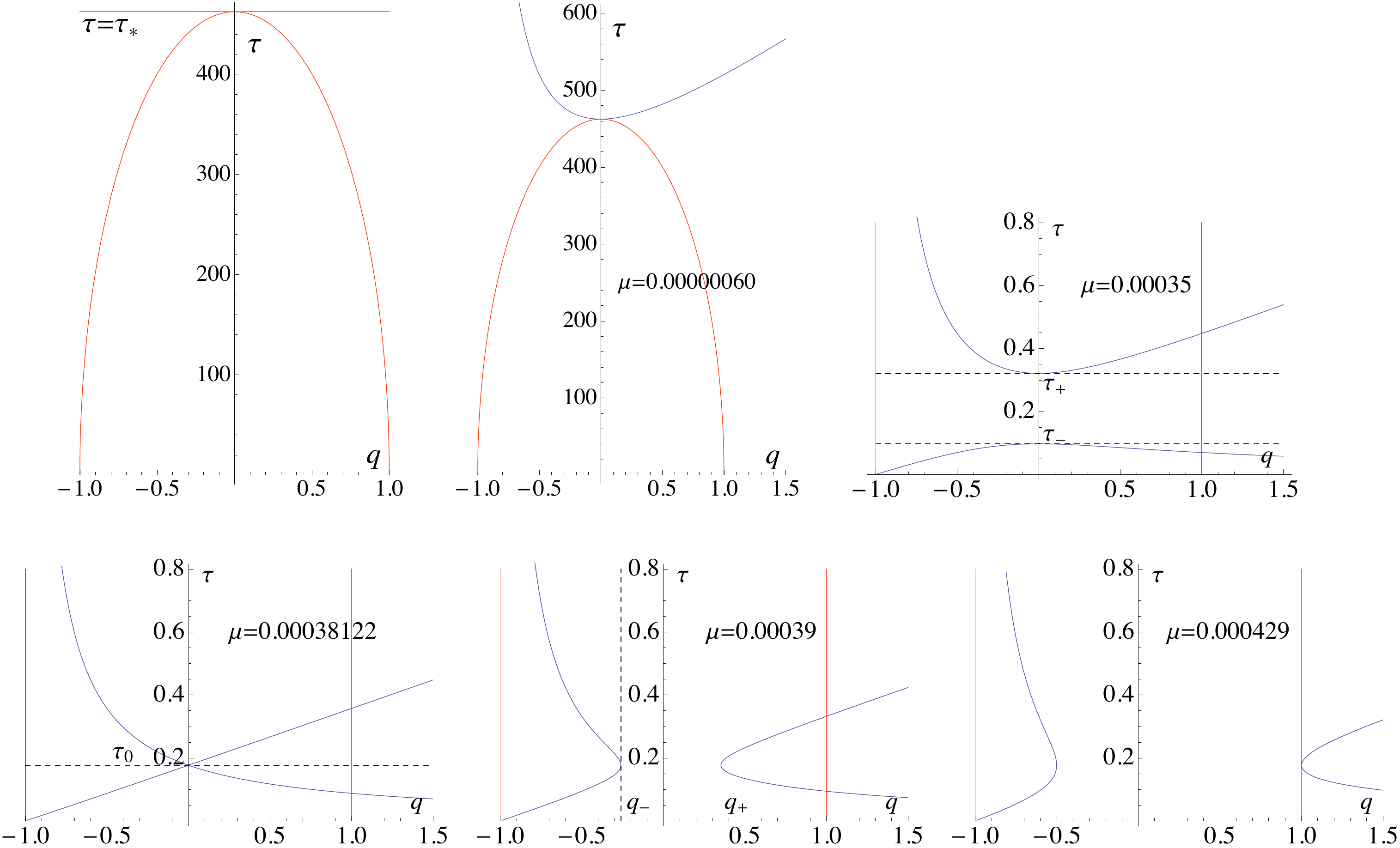}}
}
\caption{\small The relations (\ref{qtausystem}ab) for small $\gamma$ and $\mu$.  The intersections of the two relations correspond the the transition frequencies.  Here, $\gamma=0.1$ and $E_0=4$, and $\mu$ increases through values ranging from the $\omega_1$-$\omega_2$ bifurcation at $\mu\sim\gamma^4/C_{12}\approx\gamma^4/166\approx6.0\text{e-7}$, through the crossing at $\mu\sim\gamma^2/C\approx\gamma^2/26.2\approx0.00038122$, to the $\omega_1$-$\omega_2$ bifurcation at $\mu\sim\gamma^2/C_{23}\approx\gamma^2/23.3\approx0.000429$.}
\label{fig:maindiagram}
\end{figure}

These two algebraic relations between $q$ and $\tau$ are shown in Figs.~\ref{fig:Graph} and \ref{fig:maindiagram}; the symmetric one is (\ref{qtausystem}a).  They provide a transparent graphical means of analyzing the transition frequencies and their bifurcations.  The $\tau$-values of the points of intersection between the two relations determine these frequencies through $\omega=E_1+\tau^{-1}$.  It is visually clear that there are at most three intersection points, and we give a proof of this in section~\ref{subsec:proof}.  Fig.~\ref{fig:Graph} shows the evolution of the relations as $\mu$ increases, for a fixed value of $\gamma$.  Initially, there is a single intersection, corresponding to $\omega_3$.  At a critical value of $\mu$, another intersection appears at the top, which then splits into two intersections corresponding to $\omega_1$ ($q<0$) and $\omega_2$ ($q>0$).  This is the $\omega_1$-$\omega_2$ bifurcation.  After a mountain-pass mutation of relation (\ref{qtausystem}b), the intersection points corresponding to $\omega_2$ and $\omega_3$ approach each other, fuse together, and then disappear on the lower right half of the relation (\ref{qtausystem}a) in the $\omega_2$-$\omega_3$ bifurcation.

When $\gamma$ tends to zero, the peak of relation (\ref{qtausystem}a) grows without bound.  But the frequency of the $\omega_2$-$\omega_3$ bifurcation remains of order 1 and (\ref{qtausystem}a) appears as two practically vertical lines, one at $q=-1$ and one at $q=1$.  This regime is depicted in Fig.~\ref{fig:maindiagram}.

\begin{figure}
\centerline{
\scalebox{0.55}{\includegraphics{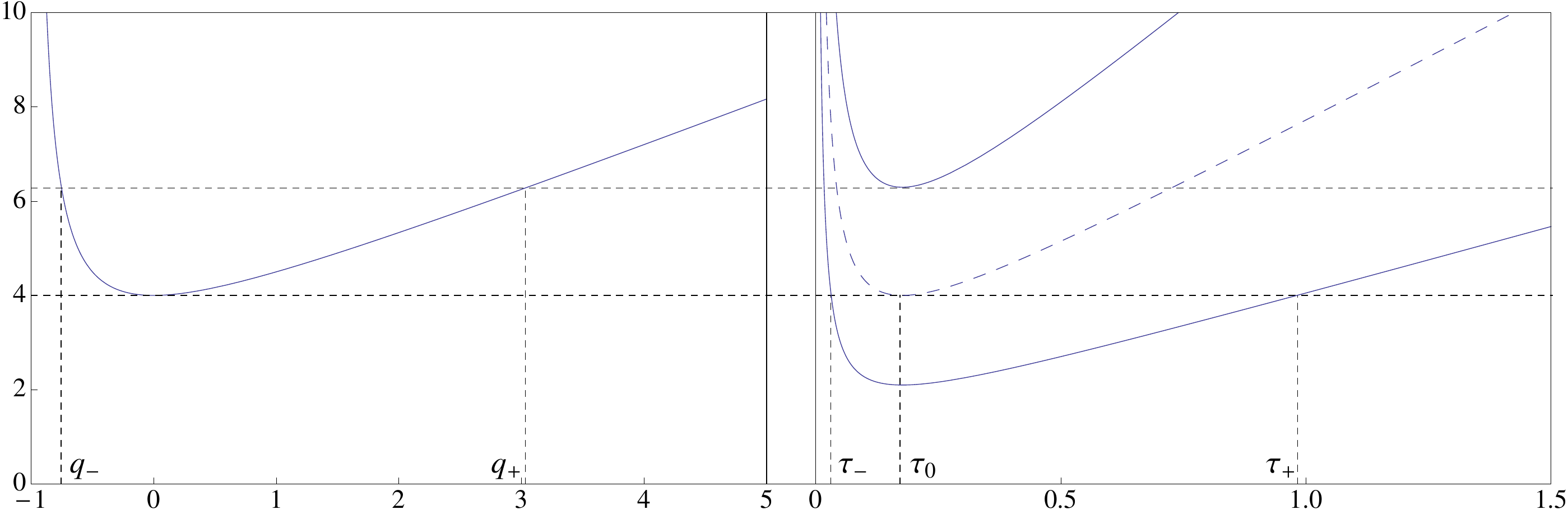}}
}
\caption{\small The concave functions $f_1(q)$ and $f_2(\tau)$ in the relation (\ref{qtausystem}b).  The function $f_1(q)$ does not depend on the parameters of the system, whereas $f_2(\tau)$ does.  When the minimal value of $f_2$ is less than that of $f_1$ (equal to $4$), the relation $f_1(q)=f_2(\tau)$ possesses two components that are functions of $q$, with minimal and maximal $\tau$-values equal to $\tau_-$ and $\tau_+$, as in the first three graphs of Fig.~\ref{fig:Graph}.  When $\min f_2(\tau)>4$, the two components are functions of $\tau$, with minimal and maximal $\tau$-values equal to $q_-$ and $q_+$ as in the last three graphs of Fig.~\ref{fig:Graph}.}
\label{fig:tauvsq}
\end{figure}

The graphical realizations of relations (\ref{qtausystem}) in Fig.s~\ref{fig:Graph} and \ref{fig:maindiagram} are obtained as follows.  In (\ref{qtausystem}a), the function of $\tau$ maps the positive real line onto itself in a strictly increasing manner.
Thus $\tau$ is implicitly a function of $q\in(0,1)$ with infinite slope at $q=\pm1$, and we denote this symmetric function by $\tau={\cal T}(q)$.

Relation (\ref{qtausystem}b) can be understood by placing the graphs of the functions
\begin{eqnarray}\label{twofunctions}
  && f_1(q) = \frac{(2+q)^2}{1+q}, \\
  && f_2(\tau) = \displaystyle \frac{9}{2}\frac{\mu}{\gamma^2}\frac{(1+a\tau)^2(1+E_1\tau)}{\tau(1+(a+\epsilon)\tau)},
\end{eqnarray}
side by side, as shown in Figure~\ref{fig:tauvsq}.  The function $f_1$ is convex and tends to infinity as $q\to-1$ or $q\to\infty$; its minimal value of $4$ is achieved at $q=0$.  Likewise, $f_1$ is convex and tends to infinity as $\tau\to0$ or $\tau\to\infty$, and thus it has a minimal value, say $m_0$.  If $m_0<4$, then (\ref{qtausystem}b), or $f_1(q)=f_2(\tau)$, has two components, each of which is the graph of a function of $q$, one with unique local maximum $\tau_-$ and the other with unique local minimum $\tau_+$, with $\tau_-<\tau_+$, both of which are achieved at $q=0$.  If $m_0>4$, then each of two components is the graph of a function of~$\tau$, the lower having maximal value $q_-$ and the upper having minimal value $q_+$, with $q_-<0<q_+$.  At the transition from one regime to the other, when $m_0=4$, the relation consists of two curves crossing tangentially at $q=0$.

If $\tau$ is viewed as a multi-valued function of $q$, then the upper branch of (\ref{qtausystem}b) is decreasing for $q<0$ and increasing for $q>0$.  Since (\ref{qtausystem}a) has the opposite behavior, it intersects the upper branch of (\ref{qtausystem}b) either not at all, exactly once at $q=0$, or at two points, one with $q<0$ and one with $q>0$.  One can see that the $\tau$-value of the former is larger by the observation that $f_1(-q)-f_1(q)=2q^3/(1-q^2)>0$, for $q\in(0,1)$.  It is clear from the graphs that the lower branch intersects (\ref{qtausystem}a) at most once, and only for $q>0$; a rigorous statement is proved in section~\ref{subsec:proof}.

\subsection{The $\omega_1$-$\omega_2$ bifurcation}\label{subsec:12bifurcation} 

We first analyze the bifurcation occurring at the point $B_{12}$ of Fig.~\ref{fig:bifurcation}, at which a narrow frequency interval of multi-valued harmonic solutions is born as $\mu$ increases across a threshold value (with $\gamma$ fixed).  We prove the power law $\gamma^4\sim C_{12}\mu$ and the asymptotics of the transition frequencies at this bifurcation.  The sharp feature in the bifurcation at the point $B_{12}$ is in fact cuspidal with $\omega_2-\omega_1 \sim C(\mu-\mu_0)^{3/2}$, meaning that the $(\omega_1,\omega_2)$ interval opens slowly.

The maximal value of the function $\tau={\cal T}(q)$ is attained at $q=0$, and we denote it by
\begin{equation*}
  \tau_* = \nu_*^{-1} = \max_{q\in(0,1)}{\cal T}(q) = {\cal T}(0).
\end{equation*}
Setting $q=0$ in \eqref{qtausystem} gives
\begin{equation*}
  12\gamma^4 = \nu_*^2(\nu_*+E_1)(\nu_*+a+\epsilon)^2.
\end{equation*}
As $\gamma\to0$ with $\nu_*>0$, we have $\nu_*\to0$, $\nu_*+E_1\to E_0$, and $\nu+a+\epsilon\to E_0+4$, and thus
\begin{equation*}
  \frac{12\gamma^4}{\nu_*^2} \to E_0(E_0+4)^2 \quad (\gamma\to0),
\end{equation*}
from which we obtain
\begin{equation}\label{taustar}
  \tau_* \sim \frac{\sqrt{E_0}(E_0+4)}{2\sqrt{3}\gamma^2} = \frac{C_*}{\gamma^2} \quad (\gamma\to0).
\end{equation}

The pair of frequencies $\omega_1$ and $\omega_2$ is born (or annihilated) when the upper branch of the second relation in \eqref{qtausystem} intersects the first relation $\tau={\cal T}(q)$ at its peak, that is, when $\tau_+=\tau_*$, as shown in the middle top graph of Figure~\ref{fig:maindiagram}.

The numbers $\tau_\pm$ are the solutions of the second relation of \eqref{qtausystem} with $q=0$.  This equation can be written as
\begin{equation*}
  \eta\frac{8}{9}\frac{\gamma^2}{\mu} \tau = (1+a\tau)(1+E_1\tau),
\end{equation*}
in which
\begin{equation*}
  \eta := \frac{1+a\tau + \epsilon\tau}{1+ a\tau} = 1+ \epsilon\frac{\tau}{1+a\tau},
  \qquad
  0<\frac{\tau}{1+a\tau} < \frac{1}{E_0+4},
\end{equation*}
so that $\eta\to1$ as $\gamma\to0$.  Thus we obtain
\begin{equation}\label{tauquadratic}
  aE_1\tau^2 + \left( a+E_1- \eta\frac{8}{9}\frac{\gamma^2}{\mu} \right)\tau + 1 = 0.
\end{equation}
Setting $\tau=\tau_*$ in this equation, with $\gamma\to0$, the asymptotic relation $\tau_*\sim C_*/\gamma^2$ yields the balance of two terms,
\begin{equation*}
  \eta\frac{8}{9}\frac{\gamma^2}{\mu}\frac{C_*}{\gamma^2} \sim aE_1\frac{C_*^2}{\gamma^4}
  \qquad (\omega_1=\omega_2, \gamma\to0),
\end{equation*}
which results in the asymptotic power law for the $\omega_1$-$\omega_2$ bifurcation
\begin{equation}\label{12bifurcation}
  \boxed{\frac{\gamma^4}{\mu} \sim \frac{3\sqrt{3}}{16} E_0^{3/2} (E_0+4)^2 =: C_{12}
  \qquad (\omega_1=\omega_2,\, \gamma\to0).}
\end{equation}

Let us denote by $\omega_{12}=\omega_1=\omega_2$ the frequency at which this bifurcation takes place.  Equation \eqref{taustar} and the definitions of $\nu$ and $\tau$ give $\omega_{12}-E_0-\epsilon\sim\gamma^2/C_*$, and this together with $\epsilon\sim\gamma^2/(4+E_0)$ yields
\begin{equation}\label{omega12}
  \boxed{\omega_{12}-E_0 \sim \gamma^2\left( \frac{1}{C_{12}} + \frac{1}{4+E_0} \right)
    = \gamma^2 \frac{2\sqrt{3}+\sqrt{E_0}}{\sqrt{E_0}(E_0+4)} = \gamma^2C'_{12}
    \qquad (\gamma\to0).}
\end{equation}

As this bifurcation takes place, the third transition frequency $\omega_3$ is very large, and one can compute its asymptotic value by finding the intersection between $\tau={\cal T}(q)$ and the lower branch of the second relation of \eqref{qtausystem}.  Let us denote this intersection by $(q_3,\tau_3)$.  The maximal $\tau$-value of the lower branch is $\tau_-$, which satisfies \eqref{tauquadratic}, and is seen to be of order ${\cal O}(\gamma^2)$, and thus the first equation of \eqref{qtausystem} gives $q=1+{\cal O}(\gamma^{16})$.  Inserting this into the second of \eqref{qtausystem}, gives
\begin{equation*}
  \eta\tau_3 = \frac{\gamma^2}{C_{12}}(1+a\tau_3)(1+E_1\tau_3),
\end{equation*}
in which $\eta\to1$ as $\gamma\to0$, or
\begin{equation*}
  1+\left( a+E_1-\frac{C_{12}}{\gamma^2}\eta \right)\tau + aE_1\tau^2 = 0,
\end{equation*}
which has a solution $\tau_3\sim \gamma^2/C_{12}$.  Finally, $\omega_3= \tau_3^{-1} + E_1$, which yields
\begin{equation}\label{omega3at12}
  \boxed{\omega_3 \sim \frac{C_{12}}{\gamma^2}
  \qquad (\omega_1=\omega_2,\, \gamma\to0).}
\end{equation}

The response $\R_{12}$ of the field at the frequency $\omega_{12}$ tends to infinity as $\gamma^{-2}$.  This can be seen by inserting the asymptotic expressions (\ref{12bifurcation},\ref{omega12}) into \eqref{RP} with $q=0$:
\begin{equation*}
  \R_{12} = \left( \omega_{12}-E_0-\frac{\gamma^2}{\omega_{12}+4} \right) \frac{2}{3\mu}
  \sim \frac{1}{\gamma^2}\frac{3}{4}E_0(E_0+4).
\end{equation*}
Similarly, the response of the field that is created or annihilated at the transition frequency $\omega_3$, that is, the field corresponding to the double root of \eqref{Polynomial}, is found to be
\begin{equation*}
  \R_3 \sim \frac{2}{3}\frac{C_{12}}{\mu^2} \sim \frac{2}{3}\frac{C_{12}^3}{\gamma^8}
    \qquad (\omega_1=\omega_2,\, \gamma\to0).
\end{equation*}

Let us why the $\omega_1$-$\omega_2$ bifurcation is a cusp.
It occurs when a convex function and a concave function intersect tangentially at their extreme values, as seen the second graph of the sequences in Figures \ref{fig:Graph} and \ref{fig:maindiagram}.  The concave function is symmetric and the convex one is not.  Up to order ${\cal O}(q^3)$, these functions can be represented by
\begin{eqnarray}
  && \tau = \tau_0 - cq^2\,, \quad r>0,\, c>0\,,\\
  && \tau = \tau_0 - r\delta + aq^2 + bq^3\,, \quad a>0\,,
\end{eqnarray}
in which $\epsilon$ is a rescaling of $\mu$.  After making the substitutions $r\delta/(a+c) \mapsto \delta$ and $d=b/(a+c)$,
the intersection points $(q,\tau)$ of these two relations satisfy
\begin{equation*}
  \delta = q^2(1+ dq)\,.
\end{equation*}
The small solutions of this equation have an expansion in powers of $\sqrt{\delta}$\,,
\begin{equation*}
  \textstyle q_{1,2} = \pm\,\delta^{1/2} - \frac{d}{2}\delta \pm \frac{5}{8}d^2\delta^{3/2}+ \cdots,
\end{equation*}
and the corresponding $\tau$-coordinates are
\begin{equation*}
  \tau_{1,2} = -c\,\delta \pm d\,\delta^{3/2}+\cdots.
\end{equation*}
The difference of these is
\begin{equation*}
  \tau_1-\tau_2 = 2d\,\delta^{3/2}.
\end{equation*}
Seeing that $\tau = 1/(\omega-E_1)$ and $\delta$ is a rescaling of $\mu$, the difference $\omega_2-\omega_1$ is of order $(\mu-\mu_0)^{3/2}$.

\subsection{Between bifurcations}\label{subsec:between} 


The structural morphosis of the relation $f_1(q)=f_2(\tau)$ occurs when the minima of $f_1$ and $f_2$ are equal and is characterized by the crossing of two curves at $(0,\tau_0)$ as depicted in Figs.~\ref{fig:tauvsq} and \ref{fig:maindiagram}.  The number $\tau_0$ is where $f_2$ attains its minimum value, say $f_2(\tau_0)=m_0$.  By writing
\begin{equation*}
  f_2(\tau) = \frac{(1+(E_0+4)\tau)(1+E_0\tau)}{\tau} \eta(\tau), \qquad
  \eta(\tau) = \frac{1+a\tau}{1+(a+\epsilon)\tau},
\end{equation*}
in which $\eta(\tau)-1$ and $\eta'(\tau)$ are both ${\cal O}(\gamma^2)$ uniformly in $\tau$, one finds that 
\begin{equation*}
  \tau_0 \sim \frac{1}{\sqrt{E_0(E_0+4)}} \qquad (\gamma\to0)
\end{equation*}
and that
\begin{equation*}
  m_0 := f_2(\tau_0) \,\sim\, 9\frac{\mu}{\gamma^2}
  \left( 2+E_0+\sqrt{E_0(E_0+4)} \right)  \qquad (\gamma\to0).
\end{equation*}
The crossing occurs when $q=0$ and $\tau=\tau_0$ simultaneously in the relation $f_1(q)=f_2(\tau)$, that is, when $4=m_0$, which yields the asymptotic power law
\begin{equation*}
  \boxed{\frac{\gamma^2}{\mu} \,\to\, \frac{9}{4}\left( 2+E_0+\sqrt{E_0(E_0+4)} \right)
  \qquad (\text{at crossing},\, \gamma\to0).}
\end{equation*}

When $m_0<4$, we have $\tau_-<\tau_0<\tau_+$, where $\tau_\pm$ are the extremes of the two branches of the relation $f_1(q)=f_2(\tau)$.  Asymptotically,
\begin{equation*}
  \tau_- \,<\, \frac{1}{\sqrt{E_0(E_0+4)}} + o(\gamma) \,<\, \tau_+
  \qquad (\gamma\to0).
\end{equation*}
When $m_0>4$, the extremes $q_\pm$ are defined through
\begin{equation*}
  \frac{(2+q_\pm)^2}{1+q_\pm} = m_0.
\end{equation*}
Since $-1<q_-<0$, we have $1<(2+q_-)^2<4$ and therefore
\begin{equation*}
  1+q_- \,<\,\,
  \frac{\gamma^2}{\mu} \left( \frac{4}{9(2+E_0+\sqrt{E_0(E_0+4)})} + o(\gamma) \right)
  \,<\, 4(1+q_-).
\end{equation*}

The point $(q_-,\tau_0)$ is the rightmost point on the left branch of the relation $f_1(q)=f_2(\tau)$, and from the diagrams in Fig.~\ref{fig:maindiagram}, this point is evidently to the right of the graph of
$\tau={\cal T}(q)$, as the latter is practically a vertical line at $q=-1$ when $\tau$ is of order $1$.  The value of $q$ that satisfies ${\cal T}(q)=\tau_0$ is obtained by setting $\tau=\tau_0$ in the first equation of \eqref{qtausystem}:
\begin{equation*}
  (1-q)(1+q) \,=\, \gamma^4\left( \frac{12}{E_0^{3/2}(E_0+4)^2(\sqrt{E_0}+\sqrt{E_0+4})^3}
    + o(\gamma) \right).
\end{equation*}
Thus $q=-1+{\cal O}(\gamma^4)$ and we obtain
\begin{equation}\label{qonredcurve}
  1+q \,<\, \gamma^4\left( \frac{6}{E_0^{3/2}(E_0+4)^2(\sqrt{E_0}+\sqrt{E_0+4})^3} 
     + o(\gamma) \right)
  \qquad ({\cal T}(q)=\tau_0,\;\gamma\to0).
\end{equation}

\subsection{The $\omega_2$-$\omega_3$ bifurcation}\label{subsec:23bifurcation} 

Because of \eqref{qonredcurve} and the symmetry of ${\cal T}$, the two intersections for $q>0$ merge when $q_+=1+{\cal O}(\gamma^4)$ and $\tau\sim\tau_0$.  This gives the frequency of the $\omega_2$-$\omega_3$ bifurcation
\begin{equation}\label{omega23}
  \boxed{\omega_{23} \,\to\, E_0 + \sqrt{E_0(E_0+4)} =: C'_{23} \qquad (\gamma\to0),}
\end{equation}
and, setting $(2+q_+)^2/(1+q_+)=m_0$, the asymptotic power law relating $\gamma$ to $\mu$,
\begin{equation}\label{23bifurcation}
  \boxed{\frac{\gamma^2}{\mu} \,\sim\, 2\left(2+E_0+\sqrt{E_0(E_0+4)}\right) =: C_{23}
  \qquad (\omega_2=\omega_3, \gamma\to0).}
\end{equation}
The response at this bifurcation is obtained from \eqref{RP},
\begin{equation*}
  {\R}_{23} \,\sim\, \frac{\sqrt{E_0(E_0+4)}}{\mu}
  = \frac{2(2+E_0)\sqrt{E_0(E_0+4)} + 2E_0(E_0+4)}{\gamma^2}.
\end{equation*}

As this bifurcation takes place, the intersection that determines $\omega_1$ has $q\to-1$ as $\gamma\to0$.  This is because, with $\gamma^2/\mu\sim C_{23}$, relation (\ref{qtausystem}b) becomes stationary as $\gamma\to0$, whereas relation (\ref{qtausystem}a) contains the scaling factor $\gamma^2$ on the right-hand side.  Using $\gamma\to0$ and $q\to-1$, together with $\gamma^2/\mu\sim C_{23}$, system \eqref{qtausystem} gives
\begin{eqnarray}\label{omega1}
  \frac{1}{1+q} &\sim& \frac{9}{2C_{23}} \frac{(1+E_0\tau)(1+(4+E_0)\tau)}{\tau}, \\
  1+q &\sim& 6\gamma^4\frac{\tau^5}{(1+E_0\tau)(1+(4+E_0)\tau)^2},
\end{eqnarray}
which together yield
\begin{equation*}
  \frac{27}{C_{23}}\gamma^4\frac{\tau^4}{(1+(4+E_0)\tau)} \to 1.
\end{equation*}
Thus $\tau\to\infty$, and we obtain
\begin{equation*}
  \tau \,\sim\, \frac{C_{23}^{1/3}}{3} \gamma^{-4/3}.
\end{equation*}
Using now $\omega_1-E_0+{\cal O}(\gamma^2) = \tau^{-1}$, we obtain
\begin{equation}\label{omega1at23}
  \boxed{\omega_1-E_0 \,\sim\, \frac{3}{C_{23}^{1/3}} \gamma^{4/3}
  \qquad (\omega_2=\omega_3,\,\gamma\to0).}
\end{equation}
The response at the transition frequency $\omega_1$, corresponding to the double root of \eqref{Polynomial}, is
${\R}_1 = (\omega_1-E_0+{\cal O}(\gamma^2))/(3\mu)$, which results in
\begin{equation*}
  {\R}_1 \,\sim\, \frac{2}{3}\frac{\,C_{23}^{2/3}}{\,\gamma^{2/3}}.
\end{equation*}

\subsection{Proof of at most three transition frequencies}\label{subsec:proof}



This section contains the proof of part (1) of Theorem~\ref{thm:bifurcation}.
It can be reformulated as follows.

\begin{prop}\label{prop:transition}
The system \eqref{qtausystem}, or, equivalently, the system \eqref{Pomegasystem}, has at most two solutions with $0\leq q<1$ ($\frac{2}{3}\leq P<1$), and if $\mu$ and $\gamma$ are sufficiently small, then the system has at most one solution with $-1<q\leq0$ ($\frac{1}{3}<P\leq\frac{2}{3}$).
\end{prop}

Using the relations $3\alpha = 1-q^2$ and $27\beta/2 = 4-3q^2-q^3$,
the system \eqref{qtausystem} can be written equivalently as
\begin{equation}\label{qnusystem}
\renewcommand{\arraystretch}{1}
\left\{
  \begin{array}{rcl}
    q^2 &=& \displaystyle \frac{12\gamma^4}{\nu^2(\nu+E_1)(\nu+(a+\epsilon))^2}\,, \\
    \vspace{-1ex} \\
    \displaystyle \frac{2q^2}{1+q} &=&
    \displaystyle 9\frac{\mu}{\gamma^2}\frac{(\nu+E_1)(\nu+a)^2}{\nu(\nu+(a+\epsilon))} - 8\,,
  \end{array}
\right.  
\end{equation}
and the first shows that $|q|<1$.

{First we deal with $0\leq q<1$}.\,
Setting $G$ and $H$ equal to the right-hand-sides of the first and second equations of \eqref{qnusystem} gives $G$ as an increasing function of $\nu$ that maps $(0,\infty)$ onto $[-\infty,1)$ and thus $H$ is a well-defined function of $G\in(-\infty,1)$.  One computes that
\begin{equation*}
  \frac{d^2 H}{dG^2} = \left(\frac{dG}{d\nu}\right)^{-2}
          \left[ \frac{d^2H}{d\nu^2} - \frac{d^2G}{d\nu^2}\frac{dH}{d\nu} \left( \frac{dG}{d\nu} \right)^{-1} \right]
\end{equation*}
and that $d^2H/d\nu^2 > 0$, $d^2G/d\nu^2<0$, and $dG/d\nu>0$ (for $\nu>0$).  Thus, whenever
$dH/d\nu>0$, $d^2H/dG^2>0$ also.  Since
\begin{equation*}
  \frac{dH}{dG} = \frac{dH}{d\nu} \left( \frac{dG}{d\nu} \right)^{-1} \quad\text{and}\quad
  \frac{dG}{d\nu}>0,
\end{equation*}
we see that $d^2H/dG^2>0$ whenever $dH/dG>0$.

Let $\nu_0$ be the positive value of $\nu$ that corresponds to $G=0$, so that $[\nu_0,\infty)$ maps onto the $G$-interval $[0,1)$.  Since $H$ is a convex function of $\nu$ that tends to infinity as $\nu\to0$ or $\nu\to\infty$, it has a unique local minimum on $[\nu_0,\infty)$ (possibly at $\nu_0$), and thus $H$ has a unique local minimum as a function of $G$ (possibly at $G=0$); denote this function by $H={\cal H}(G)$.

The expressions $G=q^2$ and $H=2q^2/(1+q)$ for $q\in[0,1)$ constitute a parameterization of the relation
$H=2G/(1+\sqrt{G})$ for $G\in[0,1)$, in which $H$ is an increasing concave function of $G$.
It remains to count the number of intersections between $H={\cal H}(G)$ and $H=2G/(1+\sqrt{G})$ on the $G$-interval $[0,1)$.  Since ${\cal H}(G)$ has a unique local minimum and is convex whenever it is increasing, it intersects $H=2G/(1+\sqrt{G})$ no more than twice.

\begin{figure} 
\centerline{
  \scalebox{0.8}{\includegraphics{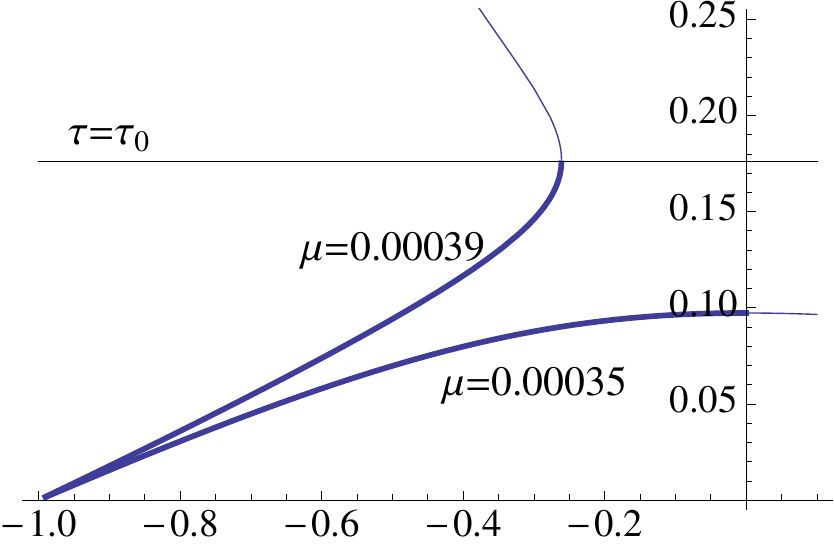}}
}
\caption{\small The function $\tau=F(q)$ in the proof in section~\ref{subsec:proof}, representing the lower branch for $q<0$ of the the relation (\ref{qtausystem}b).  It is increasing as a function of $\mu$ and convex if $\mu=\mathrm{const.}\gamma$ and $\gamma$ is sufficiently small.}\label{fig:Fofq}
\end{figure}

Now let us consider $-1<q\leq0$.
Let $\tau=F(q)$ be defined through the relation $f_1(q)=f_2(\tau)$ for $q\in(0,q_-)$ ($q_-\leq0$) and $\tau<\tau_0$.  We shall show that, for $\gamma$ sufficiently small and $\mu=\mathrm{const.}\gamma$, we have $F''(q)>0$, $1+q_->\mathrm{const.}\gamma$, and ${\cal T}(q)=\tau_0\implies 1+q = {\cal O}(\gamma^4)$.  From this, it follows that the relation $f_1(q)=f_2(\tau)$ intersects $\tau={\cal T}(q)$ exactly once for $q<0$ and these values of $\gamma$ and $\mu$.  The result is extended to $\mu<{\cal O}(\gamma)$ by the observation that, as $\mu$ decreases, $q_-$ increases (until it reaches $0$) and $F(q)$ decreases, which disallows the occurrence of any additional intersections (Fig.~\ref{fig:Fofq}).

We have already found that $\tau_0\sim1/\sqrt{E_0(E_0+4)}$, that $1+q_->\mathrm{const.}\gamma^2/\mu$, and that $\tau_0={\cal T}(p)\implies1+q_-<\mathrm{const.}\gamma^4$ as $\gamma\to0$.  Define
\begin{equation*}
  g_1(q) = \frac{1}{f_1(q)}, \qquad
  g_2(\tau) = \frac{\tau}{(1+E_0\tau)(1+(E_0+4)\tau)}, 
\end{equation*}
so that $F(q)$ satisfies the asymptotic relation
\begin{equation*}
  g_1(q) \sim \frac{\gamma^2}{\mu} g_2(F(q)) \qquad (\gamma\to0).
\end{equation*}
The second derivative of $F$ satisfies asymptotically
\begin{equation*}
  F''(q) \sim \frac{1}{g_2'(\tau)}\left( g_1''(q) - \frac{\mu}{\gamma^2}\frac{g_1'(q)}{g_2'(\tau)}g_2''(\tau) \right)
  \qquad (\tau=F(q)).
\end{equation*}
One can verify the following for $(q,\tau)$ on the graph of $F$: $g_2'(\tau)>0$, $g_1''(q)$ is bounded from below; $g_2''(\tau)$ is negative and bounded from above; if $q_-<0$, then $g_1'(q)/g_2'(\tau)$ is positive and bounded from below (in fact this ratio reaches $\infty$ at $q=q_-$, where $g_2'(\tau)=0$).  As we have seen, we can guarantee $q_-<0$ by making $\mu=\mathrm{const.}\gamma$, and then, if $\gamma$ is sufficiently small, we obtain $F''(q)>0$.


\section{Stability of Harmonic Solutions}\label{sec:stability}

To analyze the stability of the harmonic scattering solutions of the form \eqref{scatteringsolution}, we first project the system to the resonator.  This results in an equation for $z$ alone exhibiting dissipation in the form of a delayed response coming from the coupling to the Schr\"odinger string with a point-mass defect.  We then linearize about harmonic solutions $Ze^{-i\omega t}$.  Linear stability analysis is carried out by analyzing the determinant $D(s)$ of this linear system in the Laplace-transform variable $s$ in the regime of small $\gamma$ and $\mu$.  Any zero of $D(s)$ in the right half plane indicates linear instability, whereas all zeroes being in the left half plane indicates linear stability.

\begin{figure}
\centerline{
\scalebox{0.70}{\includegraphics{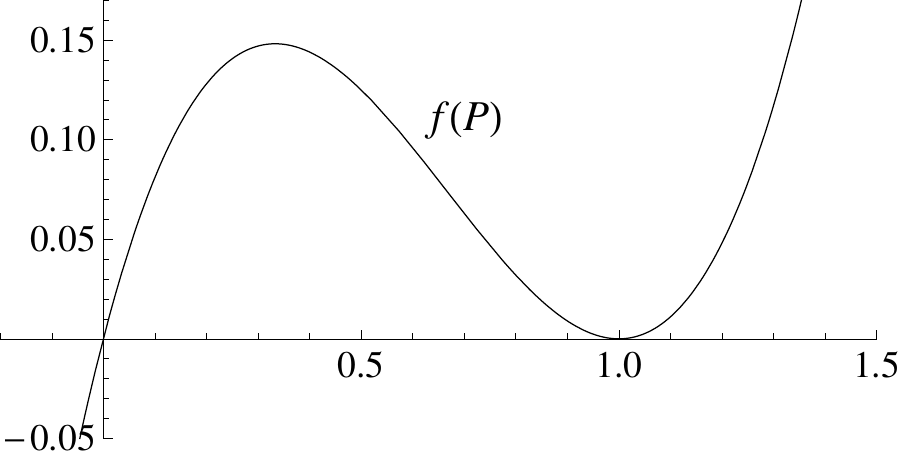}}
}\vspace{1ex}
\centerline{
\scalebox{0.222}{\includegraphics{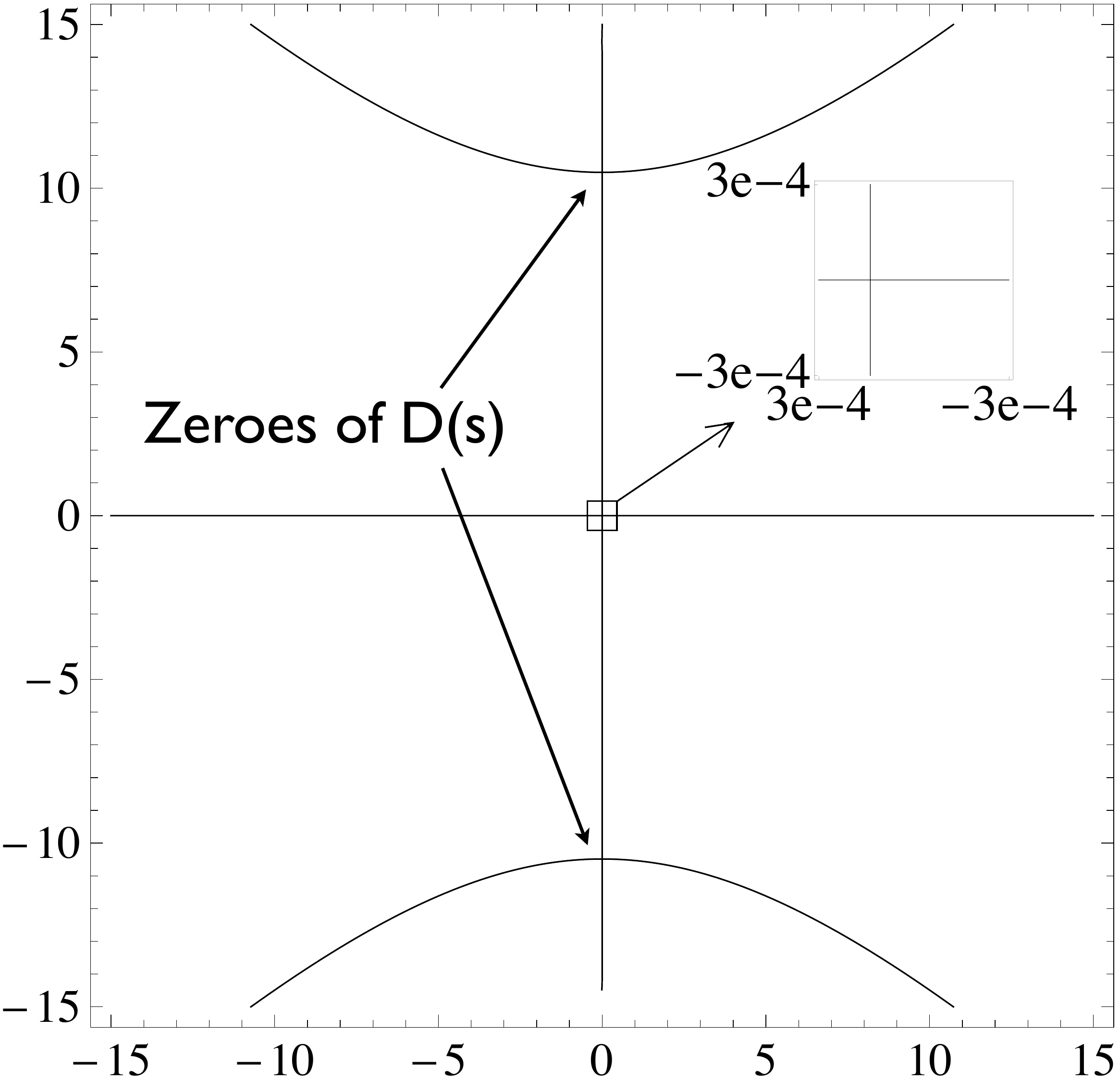}}
\scalebox{0.222}{\includegraphics{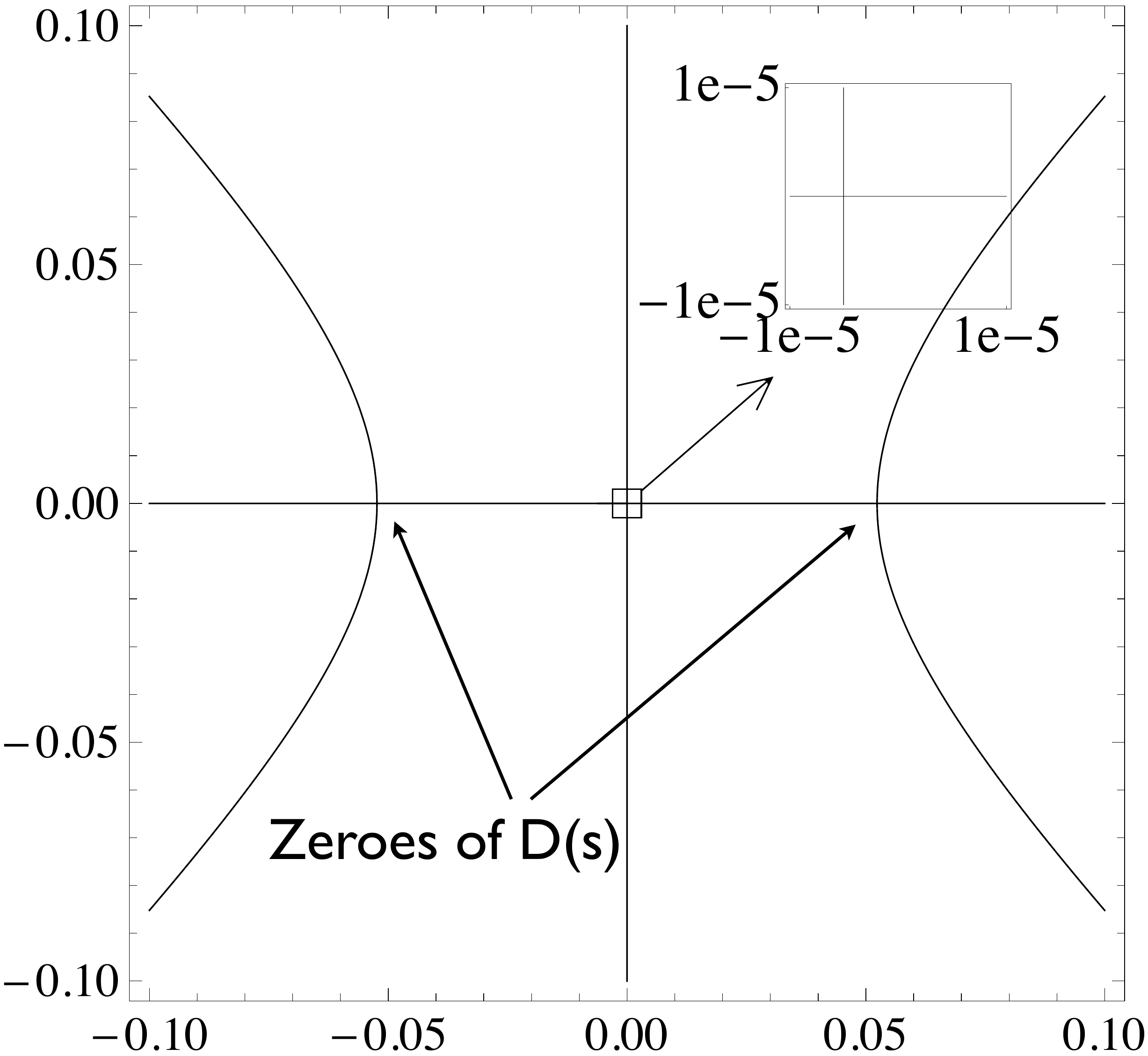}}
\scalebox{0.222}{\includegraphics{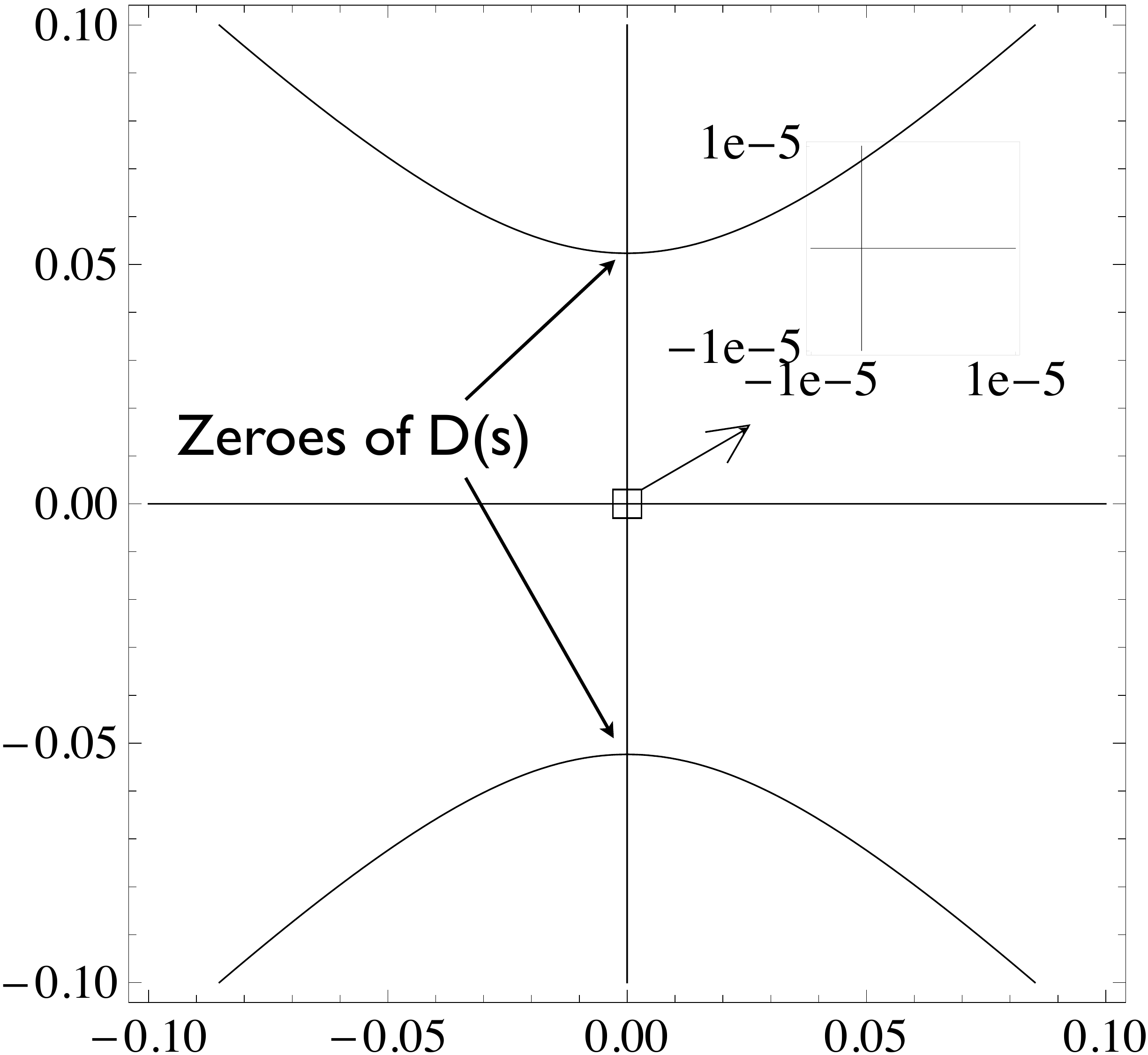}}
}\vspace{1ex}
\caption{\small {\bfseries Top.} Each of the three roots $P_i,\;i=1,2,3$ of the polynomial $f(P)=P(P-1)^2 + \alpha P - \beta$ (\ref{Polynomial}), all of which are positive, corresponds to a harmonic solution of the system (\ref{system}).  The response of resonator is related to $P$ by equation \eqref{RP}.  With the parameters $\gamma=0.05$, $\mu=4\,\gamma^2/C_{23}\approx0.000429$, and $E_0=4$, and $\omega=1.5\,\omega_{23}\approx14.5$, $P_1$ is very close to $0$, whereas $P_{2,3}$ are very close to $1$.  (There is only one transition frequency $\omega_1$ for these parameters, {\itshape i.e.,} $\mu$ is above the point $B_{23}$ in Fig.~\ref{fig:bifurcation}.)
{\bfseries Bottom.}
For each root $P_i$\,, the zero sets of the real and imaginary parts of $D(s)$ (\ref{DP}) in the complex $s$-plane are shown from left to right in increasing order of $P_i$.  {\em These graphs only show the zeros of $D(s)$ that are away from $\pm i\omega$.}
For $\gamma$ sufficiently small, the imaginary part of $D$ vanishes on the real axis and on an almost vertical curve approximately connecting the points $-i\omega$ and $i\omega$, whereas the real part vanishes on a hyperbola-like curve.  The real part of this curve is small and negative, in agreement with the asymptotic calculation in section~\ref{subsec:stability} when $\pa$ or $\pb$ is bounded.
}
\label{fig:stability}
\end{figure}

The responses $|Z/J|^2$ are obtained from the roots of the polynomial $f(P)$ (\ref{RP},\ref{Polynomial}), and it is the roots $P$ themselves that appear in the expression for $D(s)$.  When $\gamma$ and $\mu$ are small and $\omega-E_0$ is bounded from below, one root of $f$ is very close to zero, and when there are three roots, two of them are very close to 1 (Fig.~\ref{fig:stability}, top).  For the highest and lowest responses, the real parts of all zeroes of $D(s)$ are shown to have a very small negative real part asymptotically as $\gamma\to0$ and $\mu\to0$.  A precise statement in terms of $D(s)$ is made in section~\ref{subsec:stability}, and the result is the following theorem.

\begin{theorem}\label{thm:stability}

\hspace{1em}\\
\noindent
{\bfseries 1.}\ The harmonic response corresponding to the lowest root of the polynomial $f(P)$ is linearly stable for $\omega-E_0>0$ if $\gamma$ and $\mu$ are sufficiently small. \\

\vspace{-2ex}
\noindent
{\bfseries 2.}\ When the system admits three distinct harmonic scattering solutions (necessarily $\omega-E_0>0$) and if $\gamma$ and $\mu$ are sufficiently small, then \\

\vspace{-2ex}
{\bfseries a.} the solution corresponding to the middle root is linearly unstable. \\

\vspace{-2ex}
{\bfseries b.} the solution corresponding to the highest root is linearly stable if $\omega$ is large enough and linearly unstable if $E_0<\omega<1/2$.
\end{theorem}

Part (2b) has an intriguing consequence.  If $\gamma$ and $\mu$ are small enough and chosen such that the ratio $\mu/\gamma^2$ is large enough, there is a single frequency interval $(\omega_1,\infty)$ of triple harmonic solutions for which the highest response is unstable for small frequencies and stable for large frequencies.  Thus a transition from instability to stability occurs at some frequency.

\smallskip

Let us assume that, for $t\leq0$, the system is in a harmonic scattering state $(u_h(x,t),y_h(t),z_h(t))$ of the form (\ref{scatteringsolution}).  The state is perturbed for $t>0$ by forcing the resonator by a small-amplitude, temporally localized function $\varepsilon(t)$.  Make the following substitutions in the main system (\ref{system}):
\begin{eqnarray*}
  && \hspace{-3em} u(x,t) = u_h(x,t) + v(x,t), \\
  && \hspace{-3em} y(t) = y_h(t) + \eta(t), \\
  && \hspace{-3em} z(t) = z_h(t) + \zeta(t),
\end{eqnarray*}
The deviation $(v(x,t),\eta(t),\zeta(t))$ of the solution from the harmonic one vanishes at $t=0$, as does $\varepsilon(t)$, and it satisfies the system
\begin{eqnarray}
  && \hspace{-3em} iv_t + v_{xx} = 0 \quad \text{for } x\not=0,\label{perturbation:u}\\
  && \hspace{-3em} i\dot\eta = \gamma\zeta - (v_x(0^+,t) - v_x(0^-,t))\quad  \text{with }\; \eta(t) = v(0,t)\,, \label{perturbation:y}\\
  && \hspace{-3em} i\dot\zeta = E_0\zeta + \gamma\eta + \lambda\big(2|z_h|^2\zeta + z_h^2\bar\zeta 
                        + 2z_h|\zeta|^2 + \bar z_h\zeta^2 + |\zeta|^2\zeta\big) + \varepsilon(t).\label{perturbation:z}
\end{eqnarray}
In addition, we impose on $v$ an outgoing condition as $|x|\to\infty$, discussed below; see~\eqref{outgoing}.
This outgoing condition postulates decay of the $v(x,t)$ as $|x|\to\infty$.  It is symmetric in $x$ because (i) $v$~starts at rest ($v(x,0)\equiv0$), that is, $u$ is a pure harmonic solution for $t<0$ and (ii) the value $v(0,t)$ together with the requirement of decay determines $v(x,t)$ for both $x\to\infty$ and $x\to-\infty$.

\subsection{The outgoing condition}\label{subsec:outgoing}

The outgoing condition is understood through consideration of an auxiliary problem on the half-line $x\geq0$ without forcing and with a free endpoint at $x=0$,
\begin{equation}\label{auxproblem}
  \renewcommand{\arraystretch}{1}
\left.
  \begin{array}{lll}
    i\dot a = - a_{xx}, & x>0,\; t>0, & \\
    a(x,0) = 0, & x\geq0, & \text{(initially at rest)}\\
    a(x,t)\to0 & \text{as } x\to\infty\,,\; t\geq 0. & \text{(decaying at $\infty$)}
  \end{array}
\right.
\end{equation}
The Laplace transform of this system is $is\hat a=-\hat a_{xx}$ with $\hat a(s,x)\to0$ as $x\to\infty$.
The solution satisfies $\hat a_x = i^{3/2}\sqrt{s\,}\,\hat a$\,, with $\arg(i^{3/2}) = 3\pi/4$, branch cut of $\sqrt{}$ on the negative half line, and $\Re\sqrt{s}>0$, or
\begin{equation}\label{outgoingright}
  \hat a(x,s) = \hat a(0,s)\,e^{ i^{3/2}\sqrt{s\,}\,x}\,. \quad (x\geq0)
\end{equation}
The inverse Laplace transform gives the general solution in terms of the Laplace-transformed value of $a(0,t)$,
\begin{equation*}
  a(x,t) = \frac{1}{2\pi i} \int\limits_{-i\infty+0}^{i\infty+0} \hat a(0,s)\,e^{ i^{3/2}\sqrt{s\,}\,x} e^{st} ds.
\end{equation*}
The part of the integral along $s=-i\omega+0$ ($\omega>0$) is a superposition of radiating (outward traveling) waves, and the part along $s=i\omega+0$ ($\omega>0$) is a superposition of spatially evanescent fields.

An analogous argument gives the outgoing condition for a function $b(x,t)$ defined for $x\leq0$:
\begin{equation}\label{outgoingleft}
  \hat b(x,s) = \hat b(0,s)\,e^{ -i^{3/2}\sqrt{s\,}\,x}\,, \quad (x\leq0)
\end{equation}
which is equivalently expressed as $\hat b_x = -i^{3/2}\sqrt{s\,}\,\hat b$\,.

The value of $v(0,t)$ connects $v(x,t)$ to the left of the defect ($x<0$) continuously with $v(x,t)$ to the right of the defect ($x>0$), that is, one puts $\hat a(0,s) = \hat v(0,s) = \hat b(0,s)$ in \eqref{outgoingright} and \eqref{outgoingleft}.
Thus the outgoing condition for the perturbation $v$ is expressed in the Laplace variable by
\begin{equation}\label{outgoing}
  \hat v_x(x,s) =
  \renewcommand{\arraystretch}{1.2}
\left\{
  \begin{array}{ll}
    -i^{3/2}\sqrt{s\,}\, \hat v(x,s) & \text{for } x<0, \\
    i^{3/2}\sqrt{s\,}\, \hat v(x,s) & \text{for } x>0,
  \end{array}
\right.
\qquad \text{(outgoing condition)}
\end{equation}
in which \,$\arg(i^{3/2})=3\pi/4$ and $\hat v(x,s)$ is continuous in $x$.  In fact, $\hat v(x,s)$ is completely determined by $\hat v(0,s)$.

\subsection{Reduction of the system to the resonator}\label{subsec:reduction}

Because of \eqref{outgoing}, $v$ is spatially symmetric, that is, $v(x,t)=v(-x,t)$, and the jump in its derivative at $x=0$ can be expressed in a simple way:
\begin{equation}\label{if}
  v_x(0^+,t) - v_x(0^-,t) = 2v_x(0^+,t).
\end{equation}
Through the outgoing condition \eqref{outgoing}, this is expressed in the Laplace variable by
\begin{equation}\label{}
  {\cal L}\left[ v_x(0^+,t)-v_x(0^-,t) \right] = 2\,i^{3/2}\sqrt{s\,}\,\hat\eta\,.
\end{equation}
Equation \eqref{perturbation:y} now yields a relation between $\eta$ and $\zeta$,
\begin{equation}\label{etazeta}
  \eta(t) = \gamma{\cal L}^{-1}\!\left[ \hat g(s)\hat \zeta \right](t)
      = \gamma (g*\zeta)(t),
\end{equation}
in which
\begin{equation}\label{g}
  \hat g(s) = \frac{-i}{s+2\sqrt{is\,}}
\end{equation}
with the branch cut for $\sqrt{}$ on the negative real half-axis and $\sqrt{r}>0$ for $r>0$.  The relation \eqref{etazeta} allows one to project the system onto the resonator by considering equation \eqref{perturbation:z} for a single function $\zeta$,
\begin{equation}\label{zetaequation}
   i\dot\zeta \,=\, E_0\zeta + \gamma^2 (g*\zeta) + \lambda\big(2|z_h|^2\zeta + z_h^2\bar\zeta 
                        + 2z_h|\zeta|^2 + \bar z_h\zeta^2 + |\zeta|^2\zeta\big) + \varepsilon(t)\,.
\end{equation}
The real part of the function $\hat g(s)$ is positive for $s$ in the right half plane, which is a condition for power dissipation for a linear system ($\lambda=0$ here) discussed in \cite{FigotinSchenker2005}.

\subsection{Linearization about a harmonic solution}\label{subsec:linearization}

Equation \eqref{zetaequation} is linearized by eliminating the quadratic and cubic terms in $\zeta$ and replacing $\zeta(t)$ with the solution $\xi(t)$ of the resulting linear equation.  It is convenient to remove the oscillatory factor $e^{-i\omega t}$ and deal with the field
$\psi(t) = \xi(t)e^{i\omega t}$.  Keeping in mind that $z_h=Ze^{-i\omega t}$, one arrives at the following equation for $\psi$:
\begin{equation}\label{psi}
  i\dot\psi \,=\, (E_0-\omega)\psi + \gamma^2 p*\psi\, 
   + \lambda\big( 2|Z|^2\psi + Z^2\bar\psi \big) + \varepsilon(t)e^{i\omega t}\,,
\end{equation}
in which $p(t) = g(t)e^{i\omega t}$.  In the Laplace variable, this becomes
\begin{eqnarray}
  && \hspace{-3.5em} \big( is+\omega\!-\!E_0-\gamma^2\hat p-2\lambda|Z|^2 \big)\hat\psi - \lambda Z^2\hat{\overline\psi}
        = \hat\varepsilon|_{s-i\omega} \label{psihata}, \\
  && \hspace{-3.5em} \big( \!\!-\!is+\omega\!-\!E_0-\gamma^2\hat{\overline{p}}-2\lambda|Z|^2 \big)\hat{\overline\psi}
        - \!\lambda\bar Z^2\hat\psi = \hat{\overline\varepsilon}|_{s+i\omega} \label{psihatb}.
\end{eqnarray}
The second equation is obtained by conjugating the first, replacing s with $\bar s$, and then using the rule
$\hat{\bar f}(s)=\bar{\hat f}(\bar s)$ for $\Re(s)>0$.  All quantities are analytic in $s$ within their domains of definition.  The determinant of this system is
\begin{equation}\label{D}
  D(s) = 3\lambda^2|Z|^4 - 4\lambda|Z|^2\big( \omega-E_0-\gamma^2(\widehat{\text{Re}\,p})\, \big)
       + (-is+\omega-E_0-\gamma^2\hat{\bar p}) (is+\omega-E_0-\gamma^2\hat p),
\end{equation}
in which
\begin{eqnarray}
  && \hspace{-3em} \hat p(s) 
      = \frac{1}{\omega+is+2i\sqrt{\omega+is\,}}\,, \label{pa} \\
  && \hspace{-3em} \hat{\bar p}(s) 
      = \frac{1}{\omega-is-2i\sqrt{\omega-is\,}}\,. \label{pb}
\end{eqnarray}
The branch cut in the argument of the square root in the denominator of $\hat p$ is taken to be the negative imaginary axis and the branch is defined by $\sqrt{1}=1$; this imparts a branch cut in the $s$ variable along the half-line $\{ s=s_1 + i\omega, s_1\leq0 \}$ in the left half plane.  Enforcing the rule that $\hat{\bar p}(s)=\bar{\hat p}(\bar s)$ for $\Re(s)>0$ dictates that the branch cut for the argument of the square root in $\hat{\bar p}$ is the positive imaginary axis with $\sqrt{1}=1$; this imparts a branch cut in $s$ along the half-line $\{ s=s_1 - i\omega, s_1\leq0 \}$. 
Thus $D(s_1+is_2)$ has two branch cuts along the half-lines $\{ s=s_1\pm i\omega, s_1\leq0\}$.

With these stipulations of the square roots, the denominator of $\hat p$ vanishes at the single point $s=i\omega$ and the denominator of $\hat{\bar p}$ vanishes at the single point $s=-i\omega$.

\subsection{Stability analysis}\label{subsec:stability}

This section is dedicated to the proof of Theorem~\ref{thm:stability}, which is stated in the proposition below in terms of the roots of $D(s)$.

The linear stability of the system about a scattering solution $u_h$, that is, whether $\psi(t)$ grows or decays as $t\to\infty$, depends on the roots of~$D(s)$.  Any root in the right half $s$-plane indicates exponential growth, and all roots being in the left half plane indicates decay.
In terms of the quantity $\P\!=\!\mu\R/\sigma=\lambda|Z|^2/\sigma$, with $\sigma=\omega\!-\!E_0-\gamma^2/(\omega+4)$, $D$ has the form
\begin{equation}\label{DP}
  D = 3P^2-4P\frac{1}{\sigma} \left( \omega-E_0 - \frac{\gamma^2}{2}(\hat p+ \hat{\bar p}) \right)
    + \frac{1}{\sigma^2}(\omega-E_0-is-\gamma^2\hat{\bar p}) (\omega-E_0+is-\gamma^2\hat p).
\end{equation}
Thus $D(s)$ depends explicitly on the parameters $\omega$, $P$, and $\gamma^2$ (as well as $E_0$).  The value of $P$ is related to $u_h$ through the correspondence between harmonic solutions and real roots of the polynomial $f(P):=P(P-1)^2 + \alpha P - \beta$ (\ref{Polynomial}), which depends parametrically on $\gamma$, $\mu$, and $\omega$.
When $\alpha$ and $\beta$ are small, the smallest root $P_1$ is nearly zero and, in the case of three roots, the other two $P_1$ and $P_2$ are nearly 1.

\begin{prop}\label{prop:stability}
\hspace{1em}\\
\noindent
{\bfseries 1.}\ In the expression \eqref{DP} for $D$, let $P$ be equal to the smallest root of $f(P)$.  If $\omega-E_0>0$ is bounded from below and $\gamma$ and $\mu$ are sufficiently small, then all zeroes of $D(s)$ have (small) negative real part. \\

\vspace{-2ex}

\noindent
{\bfseries 2.} Suppose that $f(P)$ has three roots (necessarily $\omega>E_0$).


\smallskip
{\bfseries a.}\ In \eqref{DP}, let $P$ be set to the intermediate root.  If $\gamma$ and $\mu$ are sufficiently small, then $D(s)$ has a root with positive real part. \\

\vspace{-2ex}

{\bfseries b.}\ In \eqref{DP}, let $P$ be set to the largest root.  If the conditions
\begin{eqnarray*}
   &&  \omega>\frac{1}{2}\, \\
   && (\omega^2-1)(\omega-E_0)^2 > \omega^2\, \\
   &&  \frac{4\gamma^2\mu}{(\omega-E_0)^3(\omega+4)} \ll |1-P| \ll 1\, \\
   &&  \frac{\mu}{\gamma^2}>\frac{\omega-E_0}{\omega(\omega+4)}\,
\end{eqnarray*}
are satisfied and if $\gamma$ and $\mu$ are sufficiently small, all zeroes of $D(s)$ have (small) negative real part.  If $E_0<\omega<1/2$, then $D(s)$ has a zero with (small) positive real part.
\end{prop}

Let us simplify notation by putting
\begin{eqnarray*}
  && \rho=\omega+4\,,\quad \sigma=\omega-E_0-\gamma^2/\rho\,,\quad \\
  && \alpha=\frac{4\gamma^4}{\omega\sigma^2\rho^2}\,,\quad
  \beta=\frac{4\gamma^2\mu}{\sigma^3\rho}\,, \\
  && P = 1+Q\,.
\end{eqnarray*}

\medskip
\noindent
{\bfseries \slshape Case $P\ll 1$.} \
\smallskip

We assume that $\omega-E_0>0$ is bounded from below and let $\gamma$ and $\mu$ tend to zero.
In this regime, $\sigma\sim\omega-E_0$, so $1/\sigma={\mathcal{O}}(1)$.  The definitions \eqref{alphabeta} of $\alpha$ and $\beta$ show that these quantities vanish as $\gamma,\mu\to0$, and thus the smallest root $P_1$ of $f(P):=P(P-1)^2 + \alpha P - \beta$ is asymptotic to~$\beta$:
\begin{equation}\label{P<<1}
  P\,\sim\,\beta \,=\, \frac{4\gamma^2\mu}{\sigma^3\rho}\,=\,{\cal O}(\gamma^2\mu)\,.
\end{equation}

\medskip
\noindent
{\bfseries \slshape Case $P\ll 1$ and $\pa$, $\pb$ bounded.} \
\smallskip
The first two terms of \eqref{DP} vanish in this regime, and thus the third term also vanishes, yielding two cases,
\begin{equation}\label{cases}
  (\,\omega-E_0-is-\gamma^2\hat{\bar p}\, ) \to0
  \quad \text{or} \quad
  (\,\omega-E_0+is-\gamma^2\hat p\, ) \to0.
\end{equation}
Putting $s=s_1+is_2$ in the first case yields
\begin{eqnarray}\label{case1}
  && s_2 \sim -(\omega-E_0) \,<\, 0\,,\\
  && s_1 \sim -\gamma^2\,\Im\hat{\bar p}\,.
\end{eqnarray}
The sign of $-\Im\hat{\bar p}$ is the same as that of $-\Re(s_1+2\sqrt{\omega-is_1+s_2\,}\,)$.  This latter expression is asymptotic to $-\Re(2\sqrt{\omega+s_2})\sim-2\sqrt{E_0}$.  This quantity is negative by the declaration of the square root in the definition of $g$~(\ref{g}).  Thus $s_1$ is asymptotically negative.  A similar argument for the second of the cases \eqref{cases} shows that $s_2\sim \omega-E_0>0$ and that $s_1\sim\gamma^2\,\Im\hat p < 0$.  Thus the roots $s$ of $D(s)=0$ are in the left half plane.

\medskip
\noindent
{\bfseries \slshape Case $P\ll 1$ and $\pa$ or $\pb$ unbounded.} \
As we have mentioned at the end of section~\ref{subsec:linearization}, the denominator of $\pa$ vanishes only at $i\omega$ and that of $\pb$ only at $-i\omega$.  Thus, if one of these quantities is unbounded, the other remains bounded.  It suffices to analyze the case of $\pa$ being unbounded, as $\pb(s)=\overline{\pa(\bar s)}$ and $D(\bar s) = \overline{D(s)}$.

If $Re(s)>0$, then \,$\sqrt{\omega+is\,}$\, is in the upper half plane.
The square root in the definition \eqref{pa} of \,$\pa$\, takes arguments in the upper half plane into the first quadrant, and thus
\begin{equation}\label{squareroot}
 Re(s)>0 \implies Im\sqrt{\omega+is\,} >0\,. 
\end{equation}
We will prove that this condition is asymptotically inconsistent with $D(s)=0$.

The assumption that $\pa\to\infty$ implies $s\to i\omega$.
Applying this to the three terms of $D(s)$ in \eqref{DP} gives
\begin{multline*}
  D \,=\, \frac{48\gamma^4\mu^2}{\sigma^6\rho^2}(1+o(1))
      \,-\, \frac{16\gamma^2\mu}{\sigma^4\rho}\left( \sigma-\frac{\gamma^2}{2}\pa \right)(1+o(1))\,+\\ 
      \,+\, \frac{2\omega-E_0}{\sigma^2} \left(E_0(1+o(1))-\gamma^2\pa\right) (1+o(1))\, \\
      \,=\, \gamma^2\pa \left[ -\frac{1}{\sigma^2}(2\omega-E_0)(1+o(1))
                                          + \frac{8\gamma^2\mu}{\sigma^2\rho}(1+o(1)) \right]
       \,+\, \frac{1}{\sigma^2}(2\omega-E_0)E_0(1+o(1))\,+ \\
       \,-\, \frac{16\gamma^2\mu}{\sigma\rho}(1+o(1))
       \,+\, \frac{48\gamma^4\mu^2}{\sigma^4\rho^2}(1+o(1))\, \\
       \,=\, \gamma^2\pa \left( -\frac{1}{\sigma^2}(2\omega-E_0)(1+o(1)) \right)
       \,+\, \frac{E_0}{\sigma^2}(2\omega-E_0)(1+o(1)).
\end{multline*}
%
%
%
Setting $D=0$ provides the asymptotic relation
\begin{equation*}
  \pa \,\sim\, \frac{E_0}{\gamma^2}\,.
\end{equation*}
On the other hand, the definition \eqref{pa} of $\pa$ with $\omega+is\to0$ provides the relation
\begin{equation*}
  \pa \,\sim\, \frac{1}{2i\sqrt{\omega+is\,}}\,.
\end{equation*}
Combining these two asymptotic expressions for $\pa$ yields
\begin{equation*}
  \sqrt{\omega+is}\sim-i\,\frac{\gamma^2}{2E_0}\,,
\end{equation*}
which, in view of \eqref{squareroot}, is inconsistent with $\Re(s)>0$.
%

\medskip
\noindent
{\bfseries \slshape Case $P\,\sim\,1$.} \
\smallskip

To see the asymptotics of the other roots, write $f(P)=0$ in terms of $Q=P-1$:
\begin{equation*}
  f(P)=0 \quad\iff\quad
  Q^2 = \frac{4\gamma^4}{\sigma^3\rho}
             \left( \frac{\mu}{\gamma^2}(1+Q)^{-1} - \frac{\sigma}{\omega\rho} \right).
\end{equation*}
This implies the asymptotic
\begin{equation}\label{Q<<1}
  Q^2 \,\sim\, \frac{4\gamma^4}{\sigma^3\rho}
          \left( \frac{\mu}{\gamma^2} - \frac{\sigma}{\omega\rho} \right)
  \quad\text{when}\quad
  \frac{4\gamma^2\mu}{\sigma^3\rho} \ll |Q| \ll 1\,.
\end{equation}
Thus there are two values of order $\gamma^2$ that $Q$ can take on, one negative and one positive, under two conditions:
\begin{equation}\label{Q<<1condition}
  \frac{4\gamma^2\mu}{\sigma^3\rho} \ll |Q| \ll 1
  \quad\text{and}\quad
  \frac{\mu}{\gamma^2}>\frac{\sigma}{\omega\rho}
  \quad \implies \quad Q \sim \pm\, C \gamma^2\;\; (C>0)\,.
\end{equation}
Equality in place of "$>$" in the condition $\mu/\gamma^2>\sigma/(\omega\rho)$ is achieved asymptotically at the $\omega_1$-$\omega_2$ bifurcation.  In general, for a fixed asymptotic ratio $\mu/\gamma^2$, the inequality of satisfied if $\omega$ is large enough or $\sigma$ is small enough.

The expression \eqref{DP} for $D(s)$ can be written as
\begin{eqnarray}\label{D2}
  D & = & 2Q - \frac{2\gamma^2}{\sigma\rho} + \frac{\gamma^2}{\sigma}(\pa+\pb) + \frac{s^2}{\sigma^2} +
                \frac{is\gamma^2}{\sigma}(\pa-\pb)\, + \\
      && -\,4Q\frac{\gamma^2}{\sigma}\left( \frac{1}{\rho} - \frac{1}{2}(\pa+\pb) \right)\notag
            + \frac{\gamma^4}{\sigma^2\rho^2} - \frac{\gamma^4}{\sigma^2\rho}(\pa+\pb) + \frac{\gamma^4}{\sigma^2}\,\pa\,\pb\,.
\end{eqnarray}

\medskip
\noindent
{\bfseries \slshape Case $P\sim 1$ and $\pa$, $\pb$ bounded.} \
These assumptions imply
\begin{equation}
  D  \;=\;  2(Q+{\cal O}(\gamma^2)) - \frac{2\gamma^2}{\sigma\rho} + \frac{\gamma^2}{\sigma}(\pa+\pb) + \frac{s^2}{\sigma^2} +
                \frac{is\gamma^2}{\sigma}(\pa-\pb)\, +\, {\cal O}(\gamma^4)\,.
\end{equation}
Let us assume $Q\sim\pm C\gamma^2$ with $C>0$ from conditions \eqref{Q<<1condition}.
Setting $D$ to zero, one obtains $|s|\ll1$, and expanding $\pa$ and $\pb$ in $s$ gives
\begin{eqnarray*}
  && \pa+\pb = \frac{2}{\rho} - \frac{2s}{\rho\,\omega^{3/2}} + {\cal O}(|s|^2), \\
  && \pa-\pb = -\frac{4i}{\rho\,\omega^{1/2}} + {\cal O}(|s|)\,.
\end{eqnarray*}
With these expressions, the equation $D=0$ becomes
\begin{equation*}
  -2\sigma^2 Q\,\left(1+{\cal O}(\gamma^2)\right) =
      \left( s \,+\, 2\gamma^2\frac{\,\sigma(\omega-\half)}{\rho\,\omega^{3/2}} \right)^2 + {\cal O}(\gamma^4)\,.  
\end{equation*}
Because $Q\sim\pm C\gamma^2$ with $C>0$, the ${\cal O}(\gamma^4)$ on the right-hand side may be absorbed into the ${\cal O}(\gamma^2)$ on the left-hand side,
\begin{equation}\label{sforQ<<1}
  s = \sqrt{-2\sigma^2Q\,} - 2\gamma^2\frac{\,\sigma(\omega-\half)}{\rho\,\omega^{3/2}} + {\cal O}(\gamma^3)\,,
  \quad \sqrt{-2\sigma^2 Q\,} \sim c\,\gamma\quad (c\not=0).
\end{equation}

The negative root $Q \sim - C\,\gamma^2$ corresponds to the middle root $P_2$ of $f(P)$, or the intermediate response of the resonator depicted by the middle branch of the amplitude {\itshape vs.} frequency graph at the top of Fig.~\ref{fig:ZTPgraphs}.  Thus $D(s)$ has a zero in the right half plane for sufficiently small $\gamma$.

The positive root $Q \sim C\,\gamma^2$ corresponds to the largest root $P_3$ of $f(P)$, or the highest response of the resonator depicted by the top branch of the amplitude {\itshape vs.} frequency graph.  The first term of \eqref{sforQ<<1} is imaginary, so the second term, which is real, determines the sign of the real part of $s$.  If $E_0<1/2$, then one has $\Re(s)>0$ for $\omega<1/2$; otherwise, $\Re(s)<0$.  

\medskip
\noindent
{\bfseries \slshape Case $P\sim 1$ and $\pa$ or $\pb$ unbounded.} \
Again, it suffices to analyze the case of $\pa$ being unbounded.  
Let us suppose that $\pa$ is unbounded and $\pb$ is bounded.  The balance of dominant terms in \eqref{D2} yields
\begin{equation}\label{balance}
  \frac{\gamma^2\pa}{\sigma}(1+is) \,\sim\, -\frac{s^2}{\sigma^2}\,.
\end{equation}
Formula~\eqref{pa} with $\pa\to\infty$ gives
\begin{equation*}
  s\sim i\omega\,,
  \quad
  \pa \,\sim\, \frac{1}{2i\sqrt{\omega+is\,}}\,.
\end{equation*}
Using the second of these in \eqref{balance} gives
\begin{equation*}
  \sqrt{\omega+is\,} \sim -\frac{\gamma^2\sigma(1+is)}{2is^2}\,,
\end{equation*}
and then using $s\sim i\omega$ in the right-hand side yields
\begin{equation*}
  \sqrt{\omega+is\,} \sim \frac{i\,\gamma^2\sigma(\omega-1)}{2\omega^2}\,
\end{equation*}
as long as $\omega\not=1$.  At $\omega=1$, $(1+is)=o(1)$.  In any case, we obtain
$\omega+is={\cal O}(\gamma^4)$, so that
\begin{equation*}
  s^2 = -\omega^2 + {\cal O}(\gamma^4).
\end{equation*}
Each of the terms $is$ and $s^2$ appears only once explicitly in \eqref{D2}, and they may be replaced by $-\omega$ and $-\omega^2$ committing an error of only ${\cal O}(\gamma^4)$.  Let us also introduce the proper scaling of $Q$ from the condition \eqref{Q<<1condition}, namely $Q=\gamma^2\tilde Q$:
\begin{eqnarray}\label{D3}
  D &=& 2\gamma^2\tilde Q - \frac{2\gamma^2}{\sigma\rho} + \frac{\gamma^2}{\sigma}(\pa+\pb)
             - \frac{\omega^2}{\sigma^2} - \frac{\omega\gamma^2}{\sigma}(\pa-\pb) + \\
       && -4\,\tilde Q\left( \frac{\gamma^4}{\sigma\rho} - \frac{\gamma^4}{2\sigma}(\pa+\pb) \right)
             + \frac{\gamma^4}{\sigma^2\rho^2} - \frac{\gamma^4}{\sigma^2\rho\,}(\pa+\pb)
             + \frac{\gamma^4}{\sigma^2}\,\pa\,\pb + {\cal O}(\gamma^4)\,.
\end{eqnarray}
Passing all terms of order $\gamma^4$ into the error (recall that $\pb={\cal O}(1)$) and rearranging terms to isolate the quantity of interest $\gamma^2\pa$ yields
\begin{equation}\label{D4}
  D \,=\, \frac{\gamma^2\pa}{\sigma}
             \left[ 1-\omega+\gamma^2\left( 2\tilde Q + \frac{\rho\pb-1}{\sigma\rho} \right) \right]
          - \left[ \frac{\omega^2}{\sigma^2} - \gamma^2\left( 2\tilde Q - \frac{2}{\sigma\rho} + \frac{1}{\sigma}\pb + \frac{\omega}{\sigma}\pb \right) \right] + {\cal O}(\gamma^4)\,.
\end{equation}
Now setting $D=0$ gives
\begin{multline}\label{pa2}
  \gamma^2\pa = \\
  -\frac{\omega^2}{\sigma(\omega-1)}
   \left[ 1 - \frac{\gamma^2\sigma^2}{\omega^2} \left( 2\tilde Q - \frac{2}{\sigma\rho} + \frac{1}{\sigma}\pb + \frac{\omega}{\sigma}\pb \right) + {\cal O}(\gamma^4) \right]
   \left[ 1 + \frac{\gamma^2}{\omega-1}\left( 2\tilde Q + \frac{\rho\pb-1}{\sigma\rho} \right) + {\cal O}(\gamma^4) \right]\,.
\end{multline}
The real and imaginary parts of this quantity are
\begin{eqnarray}\label{realimag}
  && -A \,:=\, \Re \gamma^2\pa = -\frac{\omega^2}{\sigma(\omega-1)} + {\cal O}(\gamma^2)\,, \\
  && \gamma^2 B \,:=\, \Im \gamma^2\pa = \frac{\gamma^2\omega^2\,\Im\pb}{\sigma(\omega-1)}
       \left( \frac{(\omega+1)\sigma}{\omega^2} - \frac{1}{(\omega-1)\sigma} \right) + {\cal O}(\gamma^4)\,,
\end{eqnarray}
in which $\pb$ is evaluated asymptotically using $\omega+is=\gamma^4\xi$, where $\xi$ is bounded:
\begin{equation}\label{pbasymptotic}
  \pb = \frac{1}{\sqrt{2\omega-\gamma^4\xi\,}(-2i+\sqrt{2\omega-\gamma^4\xi\,})}
  \sim \frac{1}{\sqrt{2\omega}(-2i + \sqrt{2\omega})}
\end{equation}
The quantity $\gamma^2\pa$ can be evaluated alternatively directly from its definition~\eqref{pa}:
\begin{multline*}
  \pa \,=\, \frac{1}{\sqrt{\omega+is\,}(2i + \sqrt{\omega+is\,})}
        \sim \frac{-i}{2\gamma^2\sqrt{\xi}\,(1-\frac{i}{2}\gamma^2\sqrt{\xi})} \\
        \sim \frac{-i}{2\gamma^2\,\sqrt{\xi}}
               \left( 1 + \frac{i\gamma^2\sqrt{\xi}}{2} - \frac{\gamma^4\xi}{4} + \cdots \right)
        \sim \frac{-i}{2\gamma^2\sqrt{\xi}} + \frac{1}{4} + \frac{i\gamma^2\sqrt{\xi}}{8} + \cdots,
\end{multline*}
which results in the asymptotic
\begin{equation}\label{pa3}
  \gamma^2\pa \,=\, \frac{-i}{2\sqrt{\xi}} + \frac{\gamma^2}{4} + {\cal O}(\gamma^4).
\end{equation}
By equating expressions \eqref{pa2} and \eqref{pa3} for $\gamma^2\pa$, we obtain
\begin{equation*}
  \frac{i}{2\sqrt{\xi}} = A - i\gamma^2 B + \frac{\gamma^2}{4} + {\cal O}(\gamma^4) = \half(\tilde A -i\gamma^2\tilde B),
\end{equation*}
in which $\tilde A$ and $\tilde B$ are real and differ from $A$ and $B$ by order ${\cal O}(\gamma^2)$.
If $\omega$ is sufficiently large, $A$ and $B$ are positive and bounded from below in magnitude.  Specifically, it is sufficient that $(\omega^2-1)(\omega-E_0)^2 > \omega^2$.
This yields $\Im\pb>0$ asymptotically.
\begin{equation*}
  \sqrt{\xi} \,=\, \frac{-1}{2(i\tilde A+\gamma^2 \tilde B)}.
\end{equation*}
Finally,
\begin{equation*}
  is = -\omega + \gamma^4\xi = -\omega + \frac{\gamma^4}{4} \frac{(\gamma^2 \tilde B - i \tilde A)^2}{(\gamma^4\tilde B^2+\tilde A^2)^2},
\end{equation*}
which results in $\Im s \sim \omega$ and
\begin{equation*}
  \Re s \sim -\frac{B}{2A^3}\,\gamma^6\,.
\end{equation*}
This result places the zeroes of $D(s)$ asymptotically in the left half plane, regardless of the sign of~$Q$.

\section{Discussion of continuum-oscillator models}\label{sec:discussion}

Simple continuum-oscillator systems serve a vital role in elucidating fundamental principles and phenomena in physics.
Horace Lamb, interested in how disturbances in a body subside due to the transmission of energy into an infinite ambient medium devised what is now known as the Lamb model as the simplest expression of this phenomenon \cite{Lamb1900}.  He showed that, if a harmonic oscillator is attached to an infinite string whose displacement is governed by the wave equation, the oscillator obeys the equation of the usual instantaneously damped harmonic oscillator.  In other words, the energy loss in an oscillator due to instantaneous friction can be perfectly conceived as the radiation of energy into an infinite string.  The coupled system of the oscillator and the string together is a conservative extension of the lossy system consisting of the damped oscillator alone.  It is in fact the minimal conservative extension of the damped oscillator, and it is unique up to isomorphism, as shown by Figotin and Schenker~\cite{FigotinSchenker2005}.  The string acts as a system of ``hidden variables" from the point of view of an observer who is able to make measurements only of the motion of the oscillator.

The linear version ($\lambda=0$) of our model \eqref{system} is designed specifically to allow the oscillator to be completely detached from the string in the zero-coupling limit ($\gamma=0$) without severing the string.  The case $\gamma=0$ corresponds to a resonator decoupled from a system of ``hidden" variables that itself exhibits non-resonant scattering in a line by a point-mass defect resulting in a simple but nontrivial transmission coefficient.  A very small coupling parameter $\gamma\ll1$ corresponds to a small perturbation of a decoupled system with an embedded eigenvalue $E_0$ and results in a sharp resonance.  

Because our transmission line is governed by a Schr\"odinger equation, it exhibits dispersion, which is experienced by an observer in the oscillator as a delayed response, or a non-instantaneous friction, assuming that there is no forcing originating at points along the length of the string:
\begin{eqnarray*}\displaystyle
  &\displaystyle i\,\dot\zeta(t) \,=\, E_0\,\zeta(t) \,+\, \gamma^2\!\int_0^\infty\!g(t')\,\zeta(t-t')\,dt'\,+\,\varepsilon(t)\,, \quad (\lambda=0) & \\
  &\displaystyle\hat g(s) = \frac{-i}{s+2\sqrt{is}}\,.&
\end{eqnarray*}
The function $g(t)$ obeys a power-dissipation condition, described in \cite{FigotinSchenker2005}, which in the Laplace variable is expressed by the condition that $\hat g(s)$ has a positive real part when $\Re(s)>0$.

The function $\varepsilon(t)$ in the above equation is a spatially and temporally localized perturbation of a harmonic oscillation, that produces a deviation $\zeta(t)$ in the state of the oscillator.  When the frequency of oscillation vanishes, the problem becomes that of the dissipation of finite-energy disturbances of a system initially at equilibrium.  In this case, nonlinearities of a general form have been analyzed by Komech \cite{Komech1995} when the string's motion is governed by the wave equation and the resonator is attached as in the Lamb model.  The system exhibits transitions between stationary energies of the nonlinear potential in the resonator, which resemble transitions between energy states in atoms.  In Komech's model, the energy is related to the height of the string rather than a frequency of oscillation.  A positive cubic nonlinearity has only one stationary point and all disturbances decay to zero.

At nonzero frequencies, cubic nonlinearity becomes interesting and the focus of study turns to the steady-state behavior of a nonlinear scatterer subject to a monochromatic harmonic forcing originating from a source far away.  In particular, one wants to understand how these steady oscillatory motions respond to finite-energy perturbations.  Our choice of a Schr\"odinger equation for the string was based on the form of the nonlinearity $\lambda|z|^2z$ that is natural for this equation, and which admits periodic solutions that are purely harmonic and mathematically tractable.  Perturbation about the harmonic motion $Ze^{-i\omega t}$ of the oscillator results in the equation
\begin{equation*}
  i\dot\psi \,=\, (E_0-\omega)\psi + \gamma^2\!\int_0^\infty\! p(t')\,\psi(t-t')\,dt'\, 
   + \lambda\big( 2|Z|^2\psi + Z^2\bar\psi \big) + \varepsilon(t)e^{i\omega t}\,,
\end{equation*}
\eqref{psi} for the linearized perturbed oscillation $\psi(t)e^{-i\omega t}$ of the nonlinear system.
Energy loss comes from the delayed response term $\hat p(s) = \hat g(s-i\omega)$, which also satisfies the dissipation condition that $\Re(s)>0\implies\Re(\hat p(s))>0$.  The solutions $Z$ depend in a complex way on $\gamma$ and $\lambda$, and stability analysis when these parameters are small is delicate, as demonstrated in section~\ref{sec:stability}.

The idea of projecting a conservative oscillatory system onto a lossy subsystem is an insightful one and has been discussed from different points of view in the works mentioned above.  Figotin and Schenker \cite{FigotinSchenker2005} view the whole string-oscillator system as a conservative extension of a dissipative subsystem of ``observable variables" (the oscillator).  The string realizes in a structurally unique way a space of ``hidden variables" that are responsible for the loss of energy measured by an observer confined to the oscillator.  The form of the energy dissipation (as the function $g(t)$ above in the linear case) observed in the oscillator is sufficient to determine the space of hidden motions responsible for the dissipation.
These ideas are brought to bear on the lossy Maxwell system in electromagnetics by these authors and others, such as Tip~\cite{Tip1998}.
Komech describes the projected system as an irreversible description of a larger reversible one.  This point of view had been advanced previously by Keller and Bonilla~\cite{KellerBonilla1986} as an illustration of how irreversible processes may be derived from reversible ones and was motivated by the question of whether macroscopic physical processes can be deduced from classical mechanics.

To an observer of our nonlinear system from the site of the resonator, the coupling to the string is felt 
as a combination of input energy and energy loss to damping. These two energies  balance out (in the sense of time averages), when the system is in a harmonic steady state. The balance is disturbed when the system is perturbed from steady state. To express this perspective in precise terms, we  project the system onto the resonator. Write the equation for $\psi$ above
in terms of the actual perturbation $\zeta(t)=\psi(t)e^{-i\omega t}$ of a harmonic solution $z_h=Ze^{-i\omega t}$ and reinstate nonlinear terms from \eqref{zetaequation}:
\begin{multline*}
  i\dot\zeta \,=\, E_0\zeta \,+\, \gamma^2\!\int_0^\infty\!g(t')\,\zeta(t-t')\,dt'
  \,+\, \lambda\left( 2|Z|^2\zeta + Z^2e^{-2i\omega t}\bar\zeta \right) \,+\\
  +\, \lambda \left( 2Ze^{-i\omega t}|\zeta|^2 + \bar Ze^{i\omega t}\zeta^2 + |\zeta|^2\zeta\right) \,+\, \varepsilon(t).
\end{multline*}
Both the external forcing $\varepsilon$ and the  steady-state field $z_h=Ze^{-i\omega t}$, 
induced by the incident field $Je^{i(kx-\omega t)}$, 
affect the dynamics of the system.
The field $z_h=Ze^{-i\omega t}$ depends on $\gamma$, $\mu$, and $\omega$ through the roots of the cubic polynomial $f(P)$~\eqref{Polynomial}.
If $\varepsilon(t)$ is taken to be $\varepsilon_0\delta(t)$, where $\delta(t)$ is a unit impulse at $t=0$, we can consider this equation for $t>0$ with $\varepsilon=0$ and a nonzero initial condition.

Importantly, introducing external  forcing and damping through coupling to the string leads to harmonic solutions that can still be calculated explicitly.  This is not possible if  external forcing and instantaneous damping are introduced directly as in the much studied  Duffing oscillator 
\begin{equation}\label{z}
 \ddot z + a\dot z + E_0 z + b z^3 \,=\, F_0\cos(\omega t)\,.
\end{equation}
 In spite of the similarities of cubic nonlinearity, harmonic forcing and intervals of triple solutions, these solutions are only approximately periodic in the Duffing case, when the parameters $a$, $b$ and $F_0$ are small.

\FloatBarrier

\bibliography{ShipmanVenakides2010}

\end{document}